\documentclass[%
 reprint,
 amsmath,amssymb,
 aps,
]{revtex4-2}
\newcommand{\klow}{{k_{\rm low}}}
\newcommand{\khp}{{k_{{\rm high}^+}}}
\newcommand{\khpr}{{k'_{{\rm high}^+}}}
\newcommand{\khn}{{k_{{\rm high}^-}}}
\newcommand{\kcut}{k_{\rm cut}}
\newcommand{\ke}{\rangle}
\newcommand{\bra}{\langle}
\newcommand{\alphap}{{\alpha^{+}}}
\newcommand{\alphan}{{\alpha^{-}}}
\newcommand{\cov}{{\emph{cov}}}
\usepackage{array}
\usepackage{setspace}
\newcolumntype{C}[1]{>{\centering\arraybackslash}m{#1}}
\usepackage{xcolor}

\usepackage{graphicx}% Include figure files
\usepackage{dcolumn}% Align table columns on decimal point
\usepackage{bm}% bold math
\begin{document}

\preprint{APS/123-QED}

% \title{Density Functional Theory with Stochastic Fragmented Exchange}
% \title{Quadratic Scaling Exchange for DFT with Stochastic Fragmentation}
\title{Deterministic/Fragmented-Stochastic Exchange for Large Scale Hybrid DFT Calculations }
% Force line breaks with \\

% \author{Ann Author}
%  \altaffiliation[Also at ]{Physics Department, XYZ University.}%Lines break automatically or can be forced with \\

% \collaboration{MUSO Collaboration}%\noaffiliation

\author{Nadine C. Bradbury }
\author{Tucker Allen}
\author{Minh Nguyen}
\author{Daniel Neuhauser} 
\email{dxn@ucla.edu} 

\affiliation{Department of Chemistry and Biochemistry, University of California, \\ Los Angeles, California, 90095, USA}

% \collaboration{CLEO Collaboration}%\noaffiliation

\date{\today}% It is always \today, today,

\begin{abstract}
 We develop an efficient approach to evaluate range-separated exact exchange for grid or plane-wave based representations within the Generalized Kohn-Sham DFT (GKS-DFT) framework. The Coulomb kernel is fragmented in reciprocal space, and we employ a mixed deterministic-stochastic representation, retaining long wavelength (low-$k$) contributions deterministically and using a sparse (``fragmented") stochastic basis for the high-$k$ part. Coupled with a projection of the Hamiltonian onto a subspace of valence and conduction states from a prior local-DFT calculation, this method allows for the calculation of long-range exchange of large molecular systems with hundreds and potentially thousands of coupled valence states delocalized over millions of grid points. We find that even a small number of valence and conduction states is sufficient for converging the HOMO and LUMO energies of the GKS-DFT. Excellent tuning of long-range separated hybrids (RSH) is easily obtained in the method for very large systems, as exemplified here for the chlorophyll hexamer of Photosystem II with 1,320 electrons.
 
 %and obtain good agreement to other established methods such as range-separed hybrid (RSH) DFT and stochastic GW. 
 
 %Additionally, our algorithm enables the system-specific tuning of range-separation parameter $\gamma$ for very large systems. 
\end{abstract}

\maketitle

%\tableofcontents

\section{Introduction}
The introduction of hybrid exchange and long-range hybrid functionals into density functional theory (DFT) dramatically improved their accuracy.\cite{Becke1993mixing,Becke1993exact,Stephens1994B3LYP,Heyd2003, Krukau2008rsh,baer_density_2005,savin1997} These improvements, now thirty years old, enabled the rapid growth of DFT as a standard tool in the chemistry lab, with the establishment of many popular commercial and open-source software. Unfortunately, it is this key improvement in functional design, exact exchange, that limits the size of computation feasible for most researchers with a set budget of computing power and time. Traditional Hartree-Fock type exchange requires the generation of all 2-electron integrals in a given basis, scaling naively as $\mathrm{O}(N_o^4)$ for $N_o$ spatially occupied orbitals. 

The most substantial advancement in improving the computational cost of exact exchange in ab-initio DFT has come in the form of the so-called ``resolution of the identity" (RI) methods.\cite{ROI_2012} Now widely adopted, these methods reduce exact exchange to  cubic in scaling. For the entire set of 2e-integrals, $\langle pq|rs \rangle$, one expands the identity in another auxiliary basis, $\beta$, reducing a 4-center integral tensor to a product of two 3-center integral tensors, $\langle pq|rs \rangle = \sum_{\beta} \langle pq|\beta\rangle \langle\beta|rs\rangle$. Such auxiliary basis sets are optimized with density fitting.\cite{Mintmire_1982, Vahtras_1993} With this intelligent design, one can cap the number of $\beta$ to be comparable to the number of atomic orbitals needed for the calculation,\cite{Jung_2005} but without fundamental improvements, this auxiliary basis still scales with system size.

Other efforts involve the power of parallel computing, such as fragmented systems, localized auxiliary orbitals, and sparse matrix algorithms.\cite{SCDM_2015, tommaso_fragment2022} In extended systems, the sparsity of overlap integrals allows for highly optimized localized auxiliary orbitals and near linear scaling.\cite{Prentice_2020, Graf_2018}  Multi-level fragmented approaches have also recently improved scaling, especially in spatially localized cases.\cite{tommaso_fragment2022}  Modern graphical-processing units (GPUs) also contribute to unlocking larger and larger calculations with RI methods.\cite{Martinez2008GPU, Kussmann_2021}

Separately, we introduced a stochastic formalism for Hartree-Fock or long-range exchange for grid based DFT codes.\cite{neuhauser2016stochastic_exchange} In this formalism the exchange becomes a projection to a stochastic occupied orbital, which is a random linear combination of all occupied orbitals represented on a grid basis, times  a random amplitude due to the Coulomb potential.  A statistical average over multiple random vectors converges to the matrix elements of the exchange operator. In this case, each random orbital covers the entire eigenbasis of the molecule, and the number of such operators typically does not grow with system size, and occasionally shrinks due to self averaging.\cite{neuhauser2016stochastic_exchange}

In this work we employ a different strategy whereby the individual molecular states are treated deterministically.  However, the usual cost of making all the matrix elements of the Coulomb interaction is reduced by orders of magnitude (and its scaling made constant) by the fragmented-stochastic compression approach we developed in a different context, stochastic GW. \cite{vlvcek2018swiftg}  Basically,  we have shown that data over a large grid can be efficiently represented by a stochastic basis made of many small ``fragments".  Beyond a small threshold, the error does not depend on the fragment size, only on the number of fragments, so a large number of short fragments can be used to represent efficiently data on a giant grid.  

In this work, we combine the best of sparse stochastic basis compression with the resolution of the identity technique. In short, we split the Coulomb kernel for the exchange calculation to two sets  (see also Ref. \cite{SR-SRI_2020}). The first is the large interaction at few low-wavevectors (small $k$) which is treated deterministically.  The remainder, the interaction at the very many (often millions) of high $k$'s, is represented here cheaply and accurately by fragmented stochastic compression, i.e., by representing the interaction through a small number (few thousands here) of short stochastic vectors, and this number does not increase with system size.  

The second ingredient to the present deterministic/fragmented-stochastic approach is to represent the hybrid-DFT Hamiltonian in the basis of molecular orbital states (MOs) near the Fermi energy (near-gap) from local-DFT.  Specifically, we first perform a local- (or semi-local) functional DFT calculation, by any efficient basis-set or plane-wave method.  We then divide the resulting local-DFT MOs to  core, valence and conduction, as well as high virtual orbitals which are ignored. 

The core orbitals of this preliminary calculation are assumed to be a good representation of the core orbitals in the eventual hybrid calculation.  We therefore assume that the valence and conduction orbitals of the hybrid case can be expanded from the valence and conduction MOs of the local-DFT calculation.  This restriction to top valence and bottom conduction orbitals is of course routinely done in beyond-DFT methods, such as RPA, TD-DFT and the Bethe-Salpeter Equation.
% Given the limited space of valence and conduction orbitals of the local-DFT case, we represent the 2-electron integral matrix in this set, with the deterministic/stochastic fragmented basis kernel, the size of which is independent of system size.

With the introduction of sparse stochastic compression to the plane wave auxiliary basis, the scaling of the resulting approach is very gentle with system size, so that in practice the hybrid exchange correction costs less than the underlying local-DFT calculation.  Further, the approach is easily parallelizable. We label it as near-gap Hybrid DFT (ngH-DFT).

In the sections below, we develop the ngH-DFT formalism, benchmark its convergence for naphthalene and fullerene, and then show the method's power by solving for a hexamer dye complex, a large system of biological significance. The proper inclusion of exact exchange here in such a large biomolecule is promising for future use of  general post-DFT methods in giant systems.

\section{Methodology}
\subsection{Hybrid DFT in the Valence-Conduction Subspace}

We begin with the Kohn-Sham (KS) orbitals $\{\phi_s\}$ and associated eigenvalues $\{\varepsilon_s\}$ of a ground-state DFT calculation approximately satisfying $h_0\phi_s\approx\varepsilon_s\phi_s$. It is not necessary that the starting calculation be fully converged, and it can originate from LDA, PBE, or whichever DFT flavor of choice, but for simplicity it will be denoted here as LDA-DFT.  

The molecular orbitals from the LDA-DFT calculation, denoted by $\phi$, are then divided into four set of states: $N_{\rm core}$ core, $N_v(=N_o-N_{\rm core})$ valence, $N_c$ conduction, and the remainder are  high conduction states which are neglected.  

We then assume that the core states from the LDA calculation are unchanged in the GKS-DFT, i.e., 
\begin{equation}
  \psi_f=\phi_f,\,f\in{\rm core}  
\end{equation}
where $\psi$ refers to a GKS molecular orbital.  Therefore, the  $M\equiv N_v+N_c$ GKS near-gap (i.e., valence+conduction) states are assumed to be described by the valence-conduction LDA states, i.e., 
\begin{equation}
\label{eq:Cps}
    \psi_s(r)=\sum_p\phi_p(r)C_{ps},
\end{equation}
where $s,p,q$ are indices over the $M$ near-gap states.  

The converged LDA-DFT Hamiltonian is expressed as (using atomic units throughout)
\begin{equation}
    h_0= -\frac{1}{2}\nabla^2+v_{eN}^{NL}+v^0[n_0](r),
\end{equation}
with the respective terms being the kinetic energy, non-local component of electron-nucleus interaction, and the local KS potential.  The latter is a functional of the LDA density, $n_0(r)$, and contains the local electron-nucleus interaction, Hartree potential, and local exchange-correlation (XC) potential, taken here to be PW-LDA \cite{PerdewWang1992}):
\begin{equation}
    v^0[n_0](r)=v_{eN}^{local}(r)+\int{\frac{n_0(r')}{|r-r'|}dr'+v_{XC}^0[n_0]
(r)}.
\end{equation}

The electron-nucleus interaction is handled with Troullier-Martins norm-conserving pseudopotentials.\cite{TroullierMartins91} Additionally, the Martyna-Tuckerman approach is used to avoid the effect of periodic images in our simulations.\cite{MartynaTuckerman1999}  

We now turn to the GKS Hamiltonian.  Here we employ a long-range hybrid, though the same formulation applies also  to any other form, such as short-range or Becke-type fractional exchange.  Note that to avoid a cluttering of indices we write here only the closed-shell GKS formalism, but the GKS Hamiltonian would generally be spin selective (unlike the LDA-DFT).  In fact, the tuning procedure we use to yield the correct $\gamma$  requires a spin-selective Hamiltonian, as discussed later. 

The starting point is the long-range part of the Coulomb interaction, defined as $u^\gamma(|r-r'|)=\text{erf}(\gamma|r-r'|)/|r-r'|$, so for the exchange the Coulomb kernel in position space is split as \cite{savin1997}
\begin{equation}
    \frac{1}{|r-r'|} = \frac{\text{erfc}(\gamma|r-r'|)}{|r-r'|}+u^\gamma(|r-r'|).
\end{equation}
The first term dominates at short-distances and is treated locally, while the second, long-range term, is accounted for explicitly.  

Range-separated hybrid functionals excel in charge transfer and excitonic effects due to the correct $-1/|r-r'|$ asymptotic behavior of the exchange term. The use of exact exchange helps alleviate the non-physical long-range self-repulsion in the LDA potential.  The range-separation parameter $\gamma$ is best obtained by enforcing piece-wise linearity of the energy with electron number.\cite{baer_tuned_2010} 

The GKS Hamiltonian is then
\begin{equation}
    h = -\frac{1}{2}\nabla^2 + v_{eN}^{NL} + v^\gamma(r) + X^\gamma_{val}+X^\gamma_{core},
\end{equation} 
where $\gamma$ refers to one or more parameters of the hybrid exchange. The $\gamma$-dependent Kohn-Sham potential is:
\begin{equation}
    v^\gamma  (r)=v_{eN}^{local}(r) +\int{\frac{n(r')}{|r-r'|}dr'}+ v_{XC }^{{\rm SR},\gamma}[n],
\end{equation}
where ${\rm SR}$ denotes short-range and $n(r)$ is the overall density, made from a sum of core and valence densities:
\begin{equation}
    n(r)=n^{core}(r)+n^{val}(r),
\end{equation}
where  $n^{core}(r)=2\sum_{f\in {\rm core}} |\phi_f(r)|^2$.
The valence density is
\begin{equation}
    n^{val}(r) = 2\sum_i f_i|\psi_i(r)|^2 = 2\sum_{pq}\phi_p(r)P_{pq}\phi_q(r),
\end{equation}
where the density matrix is $P_{pq}  = \sum_i C_{pi}  f_i  C_{qi}$.  Here,  the sum runs over all occupied (or partially occupied) valence GKS MOs, and $f_i $ is the occupation, which can be fractional:
\begin{equation}
f_i(\varepsilon_i ;\mu)=\frac{1}{1+e^{(\varepsilon_i-\mu)/k_BT}}.
\end{equation}

The action of the valence (short-hand \emph{val}) component of the $\gamma$-dependent exact exchange operator on a general function $\eta$ is  
\begin{equation}
    (X^\gamma_{val}\eta)(r) = -\sum_i f_i  \psi^*_{i }(r)\int u^\gamma(|r-r'|)\eta(r')\psi_{i }(r')dr'.
\end{equation} %maybe include occupation # f_i? but already implicit in only summing over valence states

The contribution of the core states to the exchange part of the Hamiltonian will be done perturbatively as discussed  later.  The LDA$\to$ GKS rotation matrix, Eq.~(\ref{eq:Cps}), is initially $C_{ps}=\delta_{ps}$ and is then iterated in the SCF procedure.

The Hamiltonian matrix elements in the valence-conduction basis are      
\begin{equation}
    h_{pq}=\langle\phi_p|h_0+\delta v +X^\gamma_{val}|\phi_q\rangle,
\end{equation} where $\delta v  \equiv v^\gamma(r) - v^0(r) $ is the difference between the current GKS and initial estimate KS potentials. 

Formally, the matrix elements of the valence exact-exchange are written as a 4-index integral tensor by starting with:
\begin{equation}
    \langle\phi_q|X_{val}^\gamma|\phi_p\rangle=-\sum_if_i \langle\phi_q\psi_i |u^\gamma(|r-r'|)|\psi_i \phi_p\rangle,
\end{equation}
and inserting the expanded wavefunction gives
\begin{equation}
\label{eq:xmatrix}
    \langle\phi_q|X_{val}^\gamma|\phi_p\rangle=-\sum_{st}\langle\phi_q\phi_s|u^\gamma|\phi_t\phi_p\rangle P_{st},
\end{equation}
where real-valued orbitals are used with the chemists' convention of $\langle r r| r' r'\rangle$.

\subsection{Deterministic/Fragmented-Stochastic Representation of the Coulomb Kernel}

Our starting point is the exchange kernel in Eq. (\ref{eq:xmatrix}) which requires a generic convolution form, written schematically as
$
w(r)= \int u^\gamma(r-r') y(r') dr'.
$
This form is diagonal in reciprocal space and for finite grids it reads:
\begin{equation}
    w(k)= \frac{1}{V} \sum_k u^\gamma(k) y(k).
\end{equation}
In the Martyna-Tuckerman approach $V$ is the overall volume including full padding in each direction (i.e.,\ $V$ is $2^3=8$ times the wavefunctions volume). Further, $u^\gamma(k)$ is not necessarily positive due to the Martyna-Tuckerman construct.

Since $u^\gamma(k)$ is large at low $k$,  its action is evaluated deterministically below an assigned cutoff, $\kcut$. (The results are correct upon convergence for any $\kcut$, as this parameter only affects the speed of convergence).  Specifically, for a given $\kcut$ we  
divide $k$-space into 3 subspaces; ``low'' -- values of $k$ below $\kcut.$; ``high$^{+}$'' -- values above $\kcut$ where $u^\gamma(k)$ is positive; and ``high$^{-}$'' -- values above $\kcut$ where $u^\gamma(k)$ is negative.  The number of points in each space is denoted, respectively, as $N_\klow, \, N_\khp, $ and $N_\khn $.  Formally we write then the identity operator in the reciprocal space as 
%With the exchange kernel only depending on the distance between coordinates $r$ and $r'$ (i.e. being translationally invariant), we choose an auxiliary basis in reciprocal space ($k$). Specifically, we chose a cutoff momentum $\kcut$.

\begin{equation}
    \begin{split} 
    I = &\sum_{\klow}|\klow  \ke \bra \klow| 
 \\ + &\sum_{\khp}|\khp \ke \bra \khp|
    + \sum_{\khn}|\khn \ke \bra \khn|.
     \end{split}
\end{equation}

The Coulomb long-range operator is then
%\begin{equation}
  %  u^\gamma\left(k\right)=\sum_{\nu}\sum_{k_\nu}{\sqrt{\left|u^\gammma \left(k_\nu\right)\right|}\left|k_\nu\right\rangle g({k_\nu}) \langle k_\nu|\sqrt{\left|u^\gamma\left(k_\nu\right)\right|}},
%\end{equation}
\begin{equation}
\begin{split}
u^\gamma = &\sum_{\klow}|\klow  \ke u^\gamma(\klow) \bra \klow| \\
           &+ \sum_{\khp} \sqrt{u^\gamma(\khp)}|\khp \ke \bra \khp| \sqrt{u^\gamma(\khp)} \\
           &- \sum_{\khn} \sqrt{|u^\gamma(\khn)|}|\khn \ke \bra \khn| \sqrt{|u^\gamma(\khn)|}.  
\end{split}
\end{equation}

Next we introduce  stochastic fragmented bases \cite{vlvcek2018swiftg} for the positive and negative high-$k$ components. We detail the discussion for the high$^+$ space, and it follows analogously for the high$^-$ space. 

A set of $N_{\alphap}$ short vectors is chosen, where each is randomly positive and negative in a ``strip'', also labeled as ``fragment":
\begin{equation}
\label{eq:alphap}
    \alphap(\khp)=\pm \sqrt{\frac{N_{k^+}}{  L}}  A_{\alpha^+}(\khp).
\end{equation}
Here $A_{\alpha^+}(k)$ is a projection to a randomly placed fragment $\alpha^+$ of length $L$, i.e., is 1 within the fragment and 0 outside, so  $ \alphap(\khp)$ is randomly positive or negative in the  fragment and vanishes outside.  The strip length, $L$, is the same for each fragment. The  fragments thus randomly and uniformly sample the entire $\{|\khp\rangle$\} space.  
 
 The constant factor in Eq. (\ref{eq:alphap}) ensures that with sufficient sampling the $\alphap$ vectors form an orthonormal set, as explained below.  A technical point is that fragments that start near the edge of the $\khp$ space, i.e., that their starting point is larger than $N_\khp -  L$, need to wrap around; alternately one can zero pad the space of $N_\khp$ points by $ L$ points on both sides, and then the constant square root factor in Eq. (\ref{eq:alphap}) needs to be slightly modified.   

\begin{table*}[]
\caption{Fundamental gaps for naphthalene, fullerene, and a 476 atom hexamer dye complex. Also shown are the total number of occupied states, the maximum numbers valence and conduction states, and the range-separation parameter for each system. All energies are in eV. The atomic basis-set calculation uses the NWChem package.   Both ngH-DFT and the atomic basis-set RSH-DFT use the BNL XC functional.}
\begin{tabular}{|C{2cm}|C{0.8cm}C{0.8cm}C{0.8cm}|C{2.2 cm}|C{2cm}|C{2.5cm}|C{2.2cm}|C{2.5cm}|}
\hline
%\centering
System & $N_{o}$ & $N^{\rm max}_{v}$ & $N^{\rm max}_{c}$ & Optimal $\gamma$ ($\text{Bohr}^{-1}$) & Plane-wave LDA-DFT & Atomic Basis-Set LDA-DFT & ngH-DFT & Atomic Basis-Set RSH-DFT \\ 
\hline
Naphthalene & 24 & 24 & 104 & 0.285 & 3.34  & 3.34  & 8.63  & 8.54           \\ 
Fullerene & 120 & 120 & 480 & 0.189  & 1.63         & 1.64      & 5.42                & 5.40           \\ 
Hexamer & 660 & 200  & 400 & 0.120 & 1.23         &        & 3.81                &                  \\ \hline
\end{tabular}
\end{table*}

%For each stochastic vector of size $N_{k_{{\rm high}^\pm}}$, we pad $\{|\khp\rangle\}$ and $\{|\khn\rangle\}$ with zeros, and then do the sparse coverage in only $k_{{high}^+}$
%For each set, an array of the $k$ vectors is padded above and below with zeroes, and the sparse stochastic basis of $+1$ and $-1$'s of length $L$ randomly begins sampling the non-zero values of $k$.% 
The strip length $ L$ and the number of stochastic vectors $N_{\alphap}$ are chosen such that each $k$ point in the high$^+$ space is sufficiently ``covered'', i.e.,  will be adequately visited by the stochastic basis $\alphap$. Specifically, we choose a coverage parameter, $\cov$, that samples how often, on average, each point is sampled. The number of chosen stochastic vectors is then 
\begin{equation}
N_{\alpha^+} = \frac{\cov \cdot N_\khp}{L}. 
\end{equation}
In the limit that this coverage parameter is large the stochastic fragments form an orthonormal basis, i.e.,
\begin{equation}
 \Big\{\alphap(\khp)\alphap(\khpr)\Big\}=\delta_{\khp\, \khpr}
\end{equation}
where the large curly brackets denote a stochastic sampling with formally $\cov \to \infty $.  
In practice it is enough to use $\cov \simeq 5$.

We then define $N_{\alphap}$ states, $|\xi^+\rangle$, with components
\begin{equation}
    \langle \khp|\xi^+\rangle =\sqrt{ u^\gamma\left(\khp\right)}\ \alphap(\khp),
\end{equation}

We repeat the whole procedure for the ${\rm high}^-$ space, and end up with 
$N_{\alpha^-}$ states for the negative high-$k$ portion of the exchange kernel
\begin{equation}
    \langle \khn|\xi^-\rangle =\sqrt{\left|u^\gamma\left(\khn\right)\right|}\ \alpha^-(\khn).
\end{equation}

We now define a combined set of states, of size $N_\xi=N_{\klow}+N_{\alphap}+N_{\alpha^-}$, that is glued together via direct summation
\begin{equation}
    |\xi\rangle = \{\sqrt{\left|u^\gamma\left(\klow\right)\right|}|\klow\rangle\} \oplus \{|\xi^+\rangle\} \oplus \{|\xi^-\rangle\}.
\end{equation}
We similarly define a sign vector of length $N_\xi$
\begin{equation}
    g_\xi = \{{\rm sign}\left(u^\gamma(\klow)\right) \} 
                     \oplus \{  1 \} \oplus \{ -1 \},
\end{equation}
i.e., in addition to the sign of the interaction for the low-$k$ components, $g$ is composed of $N_\alphap$ values of $1$ and $N_\alphan$ values of $-1$.    

With these definitions, we now reach the stochastic fragmented basis representation of the exchange operator
\begin{equation}
    u^\gamma=\sum_{\xi}|\xi\rangle g_\xi\langle\xi|.
\end{equation}
This is the central equation of the deterministic/stochastic-fragment representation of the Coulomb interaction.  As mentioned, it is used here only for the exchange component and not for the direct Coulomb interaction.

\begin{figure}[]
\includegraphics{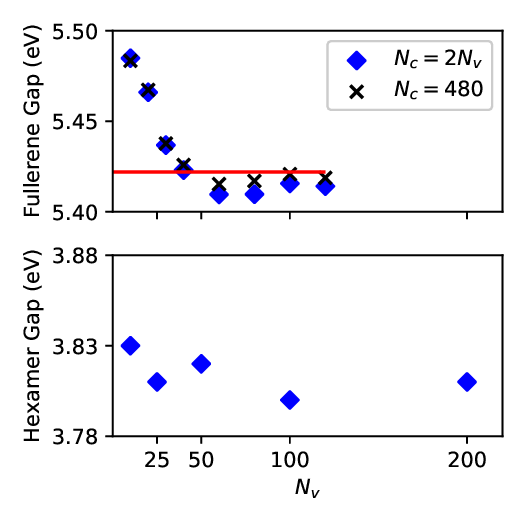}
    \caption{(Top) Convergence of the fundamental gaps of fullerene and (Bottom) hexamer with the number of valence states, $N_v$, and the number of conduction states, $N_c$, chosen either $N_c$=$2N_v$ (blue diamonds) or fixed at $N_c=480$ (black x).
     The red-line is the reference value of the fullerene gap including all occupied states, $N_v=N_o=120$ and $N_c=480$.}
\end{figure}

Inserting this form of $u^\gamma$ in the matrix element of Eq. (\ref{eq:xmatrix})
\begin{equation}
\label{eq:XvalPreU}
\langle\phi_q|X_{val}^\gamma|\phi_p\rangle=-\sum_{st\xi i}\langle\phi_q\phi_s|\xi\rangle g_\xi\langle\xi|\phi_t\phi_p\rangle C_{si}f_iC_{ti},
\end{equation}
and defining 
\begin{equation}
\label{eq:uxi}
    u_{\xi pi }\equiv\sum_t C_{ti}\langle\xi|\phi_t\phi_p\rangle,
\end{equation}
yields the final expression for the exact exchange matrix elements: 
\begin{equation}
\label{xmat_frag}
\langle\phi_q|X_{val}^\gamma|\phi_p\rangle=-\sum_{i\xi}u^*_{\xi qi }f_i g_{\xi}u_{\xi pi }.
\end{equation}
Note that for a spin-resolved calculation, the only difference is that, in addition to the amplitudes $C_{ti}$ and the exchange correlation potential $\delta v$, the transformed exchange vectors $u_{\xi p}$ and the $X_{val}^\gamma$ matrix would also gain a spin index.

\subsection{Algorithm cost}
In addition to the underlying local-DFT, the algorithm cost is  mostly  
due to preparing the $ \bra \phi_q \phi_s |\xi\ke $ and then calculating in each SCF iteration the exact exchange matrix elements. The steps are:
\begin{itemize}
    \item First one Fourier transforms, i.e., prepares $ \bra \phi_q \phi_s |k \ke $ from $\phi_q(r)\phi_s(r)$, which costs ${\mathcal{O}}(M^2 N \log N )$ operations, where $N$ is the number of total number of grid and $k$ points.  
\item Next one dot-products $\bra \phi_q \phi_s |k \ke $ with the  $N_\alpha (\equiv N_\alphap + N_\alphan)$ fragmented stochastic orbitals of length $L$ each, to yield $\bra \phi_q \phi_s | \xi^{\pm} \ke $, at a cost of $\mathcal{O}(M^2 N_\alpha L) $ operations.  For simplicity we choose here
$
N_\alpha = N_\klow = N_\xi /2.
$
Therefore, the dot product cost is $\mathcal{O}(M^2 \cdot \cov \cdot \, N_\xi )$.
\item
Finally, in each of the $N_{scf}$ iterations one prepares the matrix elements via  Eqs. (\ref{eq:uxi}) and (\ref{xmat_frag}), at a cost of $M^2 N_v N_\xi$ operations each. 
\end{itemize}

The overall cost is therefore:
\begin{equation}
\label{eq:cost}
    {\mathcal{O}}\Big({M}^2 \big( {N}(\cov +\log N)+ N_{scf} N_v N_\xi \big) \Big).
\end{equation}
Since $N_\xi$ does not grow with system size, as demonstrated below, the scaling is formally cubic with system size.  However, in practice the scaling is gentler, since a very low number of near-gap (i.e., valence+conduction) states, $M$, is sufficient for large systems.

\subsection{Core States Correction to the Exchange}
% rigid scissors of occupied/virtual evals with homo and lumo exchange energy with "core" deep state contribution
In the previous sections, the core state contributions to the exact exchange were neglected. We will account for it by a perturbative correction to the KS eigenvalues $\varepsilon_{s }\to\varepsilon_{s  }+\Delta_{s  }$,
where 
\begin{equation}
    \Delta_{s  }=\langle\psi_{s  }|X^\gamma_{core}|\psi_{s  }\rangle,
\end{equation}
is evaluated as
\begin{equation}
    \Delta_{s  }=
    -\sum_{f\in {\rm core} } 
    \bra \psi_{s   }  \phi_f 
    |u^\gamma|
   \phi_c \psi_{f   }  \ke.
\end{equation}
Since in this work we are only interested in the HOMO and LUMO energies, we calculate the correction for these two states only, labelled as $\Delta_{occ}$, $\Delta_{unocc}$.
The core-corrections stabilize the frontier orbital eigenvalues and bandgap even when the number of active valence and conduction orbitals included in the GKS-Hamiltonian is dramatically reduced.  Computationally these core corrections are very cheap as they are only added  in the last iteration, and they are calculated as explicit convolution integrals.

\section{Results}
\label{Results}
We test the ngH-DFT method with three molecular systems of increasing size: naphthalene ($N_o$=$24$), fullerene ($N_o$=$120$), and a hexamer dye complex ($N_o$=$660$). An initial PW-LDA DFT calculation is performed for all systems. The large dye system's nuclear coordinates, optimized at the PBE/def2-TZVP-MM level, were taken from \cite{Forster_ADF_Chromo_2022, psn_cnts2022}. All simulations use a generous box size that extends $6$ Bohr beyond the extent of the molecule in each direction, with real-space grids (before the Martyna-Tuckerman expansion) of $N_g$=$50,688$, $216,000$, and $2,273,280$ points respectively, and uniform grid spacings $dx$=$dy$=$dz$=$0.5$ Bohr.  The RSH-DFT studies use the Baer-Neuhauser-Livshits (BNL) XC functional.  

To balance the cost between the deterministic low-$k$ and sparse stochastic high-$k$ components of the exchange, we set, as mentioned, the size of the sparse basis, $N_{\alpha}$, equal to the number of deterministic $k$-vectors, $N_{\klow}$. The $\kcut$ parameter, separating the deterministic and fragmented-stochastic term, is adjusted so that for most of our simulations (except for a few reported in Table IV) a constant $N_{\klow}\simeq N_\alpha=5000$ is used, so the auxiliary basis size is $N_\xi\simeq 10,000$. The associated $\kcut$ values (in atomic units) are, respectively, $1.8,1.1$  and $0.5$.

Note that at these values, and  for the tuned values of $\gamma$ listed below $(0.285, 0.189$ and $0.12$ Bohr$^{-1}$, respectively), the high-$k$ interaction is very small, as $v_\gamma(k) \propto \exp(-k^2/4\gamma^2)/k^2$ (although it is numerically somewhat larger in the Martyna Tuckerman approach). For a preliminary study of the potential usefulness of the approach for other types of Hybrid functionals, where $v_\gamma(k)$ is not so tiny at high $k$, we also include later results at a lower $\kcut$.

\begin{figure}
    \centering
    \includegraphics{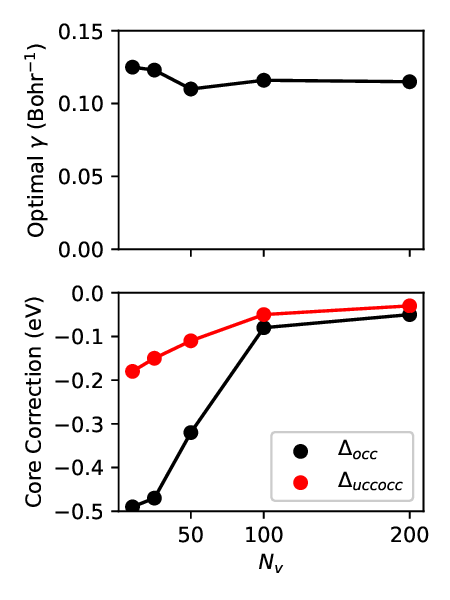}
    \caption{(Top) Convergence of $\gamma$, and (Bottom) the core corrections as a function of $ N_v$ for the hexamer system.}
    \label{fig:enter-label}
\end{figure}

Before showing the promise of using only a fraction of near-gap states, we report in Table I the fundamental gaps obtained for naphthalene, fullerene and the hexamer, using a large number of valence and conduction states (including all $N_o$ occupied states for the two smaller systems).  For naphthalene and fullerene we benchmark vs. an all-electron calculation that uses the NWChem package,\cite{NWChem2020} 
with a Gaussian aug-cc-pvdz basis  containing 302 atomic basis functions for naphthalene and 1380 for fullerene.  The fundamental gaps agree well between ngH-DFT and NWCHEM, and we demonstrate below that this agreement is maintained even when we reduce significantly the size of the valence-conduction near-gap space. 

Both the ngH-DFT and RSH-DFT calculations use the same optimal range-separation parameter $\gamma$ obtained by systematic tuning of the HOMO energies, i.e., ensuring that the HOMO energy does not change when the system is slightly ionized,  and we use here $\varepsilon_{\rm HOMO}^{\rm neutral}=\varepsilon_{\rm HOMO}^{\rm +0.1}$.  The ngH-DFT for the charged system is done via an open-shell calculation.  

A side note is that to ensure rapid convergence with the valence basis size $N_v$, we find it important to do the initial LDA calculation with the right charge, as this ensures that the core eigenstates are correctly polarized. Thus, the charged system ngH-DFT requires a initial basis-set $\phi_s$ from an LDA SCF with fractional occupation $f_{\rm HOMO}=1-0.1$ (though done in a non-spin-selective calculation) rather than relying on the $\phi_s$ from the neutral LDA.

\begin{table}[]
\caption{Naphthalene frontier orbital eigenvalues, fundamental gap, and core corrections for different numbers of valence to conduction states. All energies are in eV. 
 The first row includes all occupied states so it has no core correction.}
\begin{tabular}{|C{1.5cm}|C{1cm}C{1cm}|C{1cm}|C{1cm}C{1cm}|}
\hline
$N_v$:$N_c$ & $\varepsilon_{H}$ & $\varepsilon_{L}$ & gap & $\Delta_{occ}$ & $\Delta_{unocc}$ \\ \hline
24:104 & -8.77     & -0.14     & 8.63  &          &            \\ \hline
20:40       & -8.78     & -0.15     & 8.63  & -0.07  & -0.04    \\ \hline
10:20       & -8.72     & -0.08     & 8.64  & -0.23  & -0.03    \\ \hline
\end{tabular}
\label{nap_evals}
\end{table}

In Table II, we provide the HOMO and LUMO eigenvalues and gap for naphthalene for a chosen number of valence and conduction states. The first row in the table includes all occupied and a large number of unoccupied states, while the following two use a reduced valence-conduction space. Reduction of this active space necessitates the core corrections of the HOMO and LUMO eigenvalues. The gap is not changed much when the valence-conduction basis-set size is made smaller.

\begin{table}[]
\caption{Fullerene frontier orbital eigenvalues, fundamental gap, and core corrections for different $N_v$:$N_c$.  All energies are in eV.}
\begin{tabular}{|C{1.5cm}|C{1cm}C{1cm}|C{1cm}|C{1cm}C{1cm}|}
\hline
$N_v$:$N_c$   &  $\varepsilon_{H}$ & $\varepsilon_{L}$ & gap & $\Delta_{occ}$        & $\Delta_{unocc}$        \\ \hline
120:480 &  -8.26     & -2.84     & 5.42  &          &          \\ \hline
40:80   &  -8.20      & -2.78      & 5.42   & -0.15   & -0.12   \\ \hline
20:40   &  -8.23      & -2.76      & 5.47   & -0.42   & -0.29   \\ \hline
20:20   &  -8.23      & -2.77      & 5.46   & -0.42   & -0.29   \\ \hline
10:10   &  -8.25      & -2.83      & 5.42   & -1.12   & -0.63   \\ \hline
\end{tabular}
\label{ful_evals}
\end{table}

As Table III shows, the convergence is even better for the next bigger system, fullerene.  The number of included valence and conduction states can now be much smaller than $N_o$.   This rapid convergence with $N_v$  is also shown in Fig.1a.  The figure further shows that the results converge rapidly with the conduction basis size, so that $N_c=2N_v$  gives essentially the same result as using a very large value  of $N_c$.  

The convergence with $N_v$ further improves for the biggest system, the hexamer, as shown in Fig. 1b. The gaps shown all agree within $\pm 0.02 $~eV even for very small $N_v$ and $N_c$.   This implies that very large systems could be used with a small valence-conduction space. 

Fig. 2 shows, for the hexamer, the convergence of the range-separation parameter as well as the core corrections. The extracted $\gamma$ values are  consistent, even with a valence-conduction space of only ten valence and ten conduction orbitals. This implies that optimal tuning of long-range separated hybrids of giant systems could be done rather cheaply. 

The single-run stochastic error, i.e., the standard deviation of the energy, is shown in Table IV.  It is estimated from the results of ten independent runs. As mentioned, for $N_\klow\simeq5000$, $\kcut$ is large for each of the three studied systems so that that the values of $v_\gamma(k)$ are very small for the stochastically-sampled high-$k$ spaces.  We therefore also include results with a smaller $\kcut$ so $N_\klow\simeq500$, for $N_\alpha=500$ and $N_\alpha=5000$ (i.e., $N_\xi \simeq 1000, 5500$).  As shown, the statistical error is still quite small, about 0.01-0.03eV, and is lower than or similar to the low stochastic error associated with using a small value of  $N_v$. 

To conclude the results section, we show in Fig. 3, for the hexamer, the number of CPU-core hours needed in ngH-DFT vs. $N_v$, using standard AMD Rome processors.  The ngH-DFT cost is very small, and even for the largest sample studied $N_v=200, N_c=400$ the required effort is less  than for the underlying LDA-DFT stage.

\begin{table}[]
\caption{Fundamental gap and its standard deviation, $\sigma$, for three test systems (in eV), for different numbers of deterministic low-$k$ terms, $N_{k_{low}}$, and sizes of the sparse stochastic basis, $N_{\alpha}$.}
\begin{tabular}{|C{2cm}|C{0.5cm}C{0.5cm}|C{0.8cm}C{0.8cm}|C{1.5cm}|C{1.5cm}|}
\hline
System & $N_{v}$ & $N_{c}$ & $N_{k_{low}}$ & $N_{\alpha}$ & gap & $\sigma$ \\ \hline
Naphthalene & 20 & 40 & 501 & 500 & 8.6329 & 0.0122           \\ 
 & & & 501 & 5000 & 8.6373 & 0.0077 
\\
& & & 4987 & 5000 & 8.6344 & 0.0004
\\ \hline
Fullerene & 40 & 80 & 515 & 500 & 5.4209 & 0.0066      \\ 
 & & & 515 & 5000 & 5.4226 & 0.0051 
\\
& & & 4945 & 5000 & 5.4228  & 0.0001
\\ \hline
Hexamer & 40 & 80 & 503 & 500 & 3.7914 & 0.0286        \\ 
 & & & 503 & 5000 & 3.8018 & 0.0152 
\\
& & & 4785 & 5000 & 3.8032  & 0.0002  
\\ \hline
\end{tabular}
\end{table}

\section{Discussion}

We developed and demonstrated here a new method,  ngH-DFT, for incorporating exact exchange within a GKS-DFT framework.  Long wavelength (low $k$) components of the exchange are evaluated deterministically, and high momenta are represented by a sparse stochastic basis.  Using an underlying MO basis from a preliminary LDA calculation the frontier eigenvalues converge with a small number of included valence and conduction orbitals. 

We reiterate that this method only has stochasticity in its handling the high momenta components of the exchange, which are not  as physically important as the low components. Treating less relevant degrees of freedom stochastically works very well here when combined with the sparse compression technique.  

\begin{figure}[]
    \includegraphics{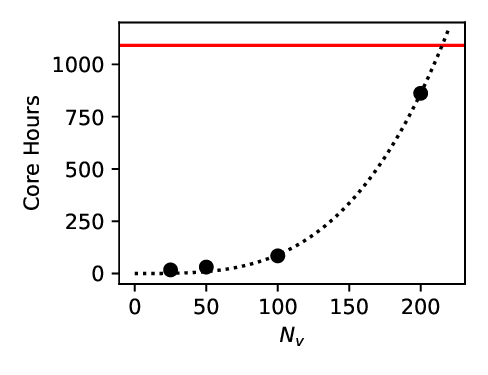}
    \caption{CPU-core hours required for the ngH-DFT method for the large hexamer complex. Parabolic scaling with the number of valence states (for a given grid) is shown. The red line indicates the core hours required for the initial LDA DFT calculation.}
\end{figure}

Future work will expand the method in several directions:  

First, the stochastic compression gave equal weight to all high-$k$ components, and could be replaced by preferred sampling of points with relatively higher $u^\gamma(k)$ within the ${\rm high}^\pm$ spaces, either explicitly or division to several sub-spaces.

Next, a relatively simple extension would be to construct random combinations of the core states that would be used to calculate the core-exchange.  This would reduce the memory requirements since the full set of core states would not need to be stored.\cite{neuhauser2016stochastic_exchange} 
Further, for the corrections of other states we could use a rigid scissor approximation \cite{Vlek2018scissors}, where the all occupied and unoccupied subspaces are shifted by the respective HOMO and LUMO orbital expectation values of $X^\gamma_{core}$; or, better yet, sample a few more states to determine an energy-dependent core-state contribution, analogous to our GW matrix elements.\cite{neuhauser2014breaking,vlvcek2019nonmonotonic}
Since it will be applied only to the core states the contribution would be small and therefore so will its underlying stochastic error.

The present near-gap approach method will be useful for many-body perturbation theory (MBPT). In MBPT methods, having access to exact exchange corrected eigenstates gives an improved starting point for methods such as one-shot $G_0W_0$ where the quality of the beginning canonical states is very important.\cite{Bruneval_2012, McKeon2022}  

Our formalism will also apply to time-dependent Hybrid-DFT, where, like in GKS-DFT SCF, the $\bra \phi_q \phi_s | \xi \ke $ vectors would be evaluated once while the exchange matrix, Eqs. (\ref{eq:uxi}), (\ref{eq:XvalPreU}) will be updated repeatedly, here once per time step.  It will be useful both for real-time TDDFT and for frequency resolved TDDFT and BSE.\cite{bradbury_bse_2022,bradbury_bse_2023} We also expect applications within basis set based DFT codes, where the wavefunction is eventually represented on a complete grid. Additionally, we anticipate that this method will have applications in auxiliary field quantum Monte-Carlo methods (AFQMC), where the bulk of the computational effort also lies in evaluating exchange energy on many Slater determinants.\cite{Rom1997Neuhauser,Carlson1999AFQMC,Zhang2018AFQMC} 

Finally, the underlying LDA-DFT approach could be efficiently done with stochastic DFT, \cite{baer2013selfaveraging,neuhauser2014fragment} so very large systems could be used, with tens of thousands of electrons or more.  Eigenstates are not produced automatically in stochastic DFT, so the set of $N_v+N_c$ near-gap eigenstates, required for ngH-DFT, would be then extracted by filter-diagonalization.\cite{Wall1995_fdg} 
\section*{Acknowledgements}
This paper was supported by the Center
for Computational Study of Excited State Phenomena
in Energy Materials (C2SEPEM), which is funded by
the U.S. Department of Energy, Office of Science, Basic
Energy Sciences, Materials Sciences and Engineering
Division via Contract No. DE-AC02- 05CH11231, as
part of the Computational Materials Sciences Program. Computational resources were provided by the National Energy Research Scientific Computing Center (NERSC), a U.S. Department of Energy Office of Science User Facility operated under Contract No. DE-AC02-05CH11231. NCB acknowledges the NSF Graduate Research Fellowship Program under grant DGE-2034835. 

\bibliography{main}% Produces the bibliography via BibTeX.

\begin{thebibliography}{40}%
\makeatletter
\providecommand \@ifxundefined [1]{%
 \@ifx{#1\undefined}
}%
\providecommand \@ifnum [1]{%
 \ifnum #1\expandafter \@firstoftwo
 \else \expandafter \@secondoftwo
 \fi
}%
\providecommand \@ifx [1]{%
 \ifx #1\expandafter \@firstoftwo
 \else \expandafter \@secondoftwo
 \fi
}%
\providecommand \natexlab [1]{#1}%
\providecommand \enquote  [1]{``#1''}%
\providecommand \bibnamefont  [1]{#1}%
\providecommand \bibfnamefont [1]{#1}%
\providecommand \citenamefont [1]{#1}%
\providecommand \href@noop [0]{\@secondoftwo}%
\providecommand \href [0]{\begingroup \@sanitize@url \@href}%
\providecommand \@href[1]{\@@startlink{#1}\@@href}%
\providecommand \@@href[1]{\endgroup#1\@@endlink}%
\providecommand \@sanitize@url [0]{\catcode `\\12\catcode `\$12\catcode
  `\&12\catcode `\#12\catcode `\^12\catcode `\_12\catcode `\%12\relax}%
\providecommand \@@startlink[1]{}%
\providecommand \@@endlink[0]{}%
\providecommand \url  [0]{\begingroup\@sanitize@url \@url }%
\providecommand \@url [1]{\endgroup\@href {#1}{\urlprefix }}%
\providecommand \urlprefix  [0]{URL }%
\providecommand \Eprint [0]{\href }%
\providecommand \doibase [0]{https://doi.org/}%
\providecommand \selectlanguage [0]{\@gobble}%
\providecommand \bibinfo  [0]{\@secondoftwo}%
\providecommand \bibfield  [0]{\@secondoftwo}%
\providecommand \translation [1]{[#1]}%
\providecommand \BibitemOpen [0]{}%
\providecommand \bibitemStop [0]{}%
\providecommand \bibitemNoStop [0]{.\EOS\space}%
\providecommand \EOS [0]{\spacefactor3000\relax}%
\providecommand \BibitemShut  [1]{\csname bibitem#1\endcsname}%
\let\auto@bib@innerbib\@empty
%</preamble>
\bibitem [{\citenamefont {Becke}(1993{\natexlab{a}})}]{Becke1993mixing}%
  \BibitemOpen
  \bibfield  {author} {\bibinfo {author} {\bibfnamefont {A.~D.}\ \bibnamefont
  {Becke}},\ }\bibfield  {title} {\bibinfo {title} {A new mixing of
  {Hartree}-{Fock} and local density-functional theories},\ }\href
  {https://doi.org/10.1063/1.464304} {\bibfield  {journal} {\bibinfo  {journal}
  {The Journal of Chemical Physics}\ }\textbf {\bibinfo {volume} {98}},\
  \bibinfo {pages} {1372} (\bibinfo {year} {1993}{\natexlab{a}})}\BibitemShut
  {NoStop}%
\bibitem [{\citenamefont {Becke}(1993{\natexlab{b}})}]{Becke1993exact}%
  \BibitemOpen
  \bibfield  {author} {\bibinfo {author} {\bibfnamefont {A.~D.}\ \bibnamefont
  {Becke}},\ }\bibfield  {title} {\bibinfo {title} {Density-functional
  thermochemistry. {III}. {The} role of exact exchange},\ }\href
  {https://doi.org/10.1063/1.464913} {\bibfield  {journal} {\bibinfo  {journal}
  {The Journal of Chemical Physics}\ }\textbf {\bibinfo {volume} {98}},\
  \bibinfo {pages} {5648} (\bibinfo {year} {1993}{\natexlab{b}})}\BibitemShut
  {NoStop}%
\bibitem [{\citenamefont {Stephens}\ \emph {et~al.}(1994)\citenamefont
  {Stephens}, \citenamefont {Devlin}, \citenamefont {Chabalowski},\ and\
  \citenamefont {Frisch}}]{Stephens1994B3LYP}%
  \BibitemOpen
  \bibfield  {author} {\bibinfo {author} {\bibfnamefont {P.~J.}\ \bibnamefont
  {Stephens}}, \bibinfo {author} {\bibfnamefont {F.~J.}\ \bibnamefont
  {Devlin}}, \bibinfo {author} {\bibfnamefont {C.~F.}\ \bibnamefont
  {Chabalowski}},\ and\ \bibinfo {author} {\bibfnamefont {M.~J.}\ \bibnamefont
  {Frisch}},\ }\bibfield  {title} {\bibinfo {title} {Ab initio calculation of
  vibrational absorption and circular dichroism spectra using density
  functional force fields},\ }\href {https://doi.org/10.1021/j100096a001}
  {\bibfield  {journal} {\bibinfo  {journal} {The Journal of Physical
  Chemistry}\ }\textbf {\bibinfo {volume} {98}},\ \bibinfo {pages} {11623}
  (\bibinfo {year} {1994})}\BibitemShut {NoStop}%
\bibitem [{\citenamefont {Heyd}\ \emph {et~al.}(2003)\citenamefont {Heyd},
  \citenamefont {Scuseria},\ and\ \citenamefont {Ernzerhof}}]{Heyd2003}%
  \BibitemOpen
  \bibfield  {author} {\bibinfo {author} {\bibfnamefont {J.}~\bibnamefont
  {Heyd}}, \bibinfo {author} {\bibfnamefont {G.~E.}\ \bibnamefont {Scuseria}},\
  and\ \bibinfo {author} {\bibfnamefont {M.}~\bibnamefont {Ernzerhof}},\
  }\bibfield  {title} {\bibinfo {title} {Hybrid functionals based on a screened
  {Coulomb} potential},\ }\href {https://doi.org/10.1063/1.1564060} {\bibfield
  {journal} {\bibinfo  {journal} {The Journal of Chemical Physics}\ }\textbf
  {\bibinfo {volume} {118}},\ \bibinfo {pages} {8207} (\bibinfo {year}
  {2003})}\BibitemShut {NoStop}%
\bibitem [{\citenamefont {Krukau}\ \emph {et~al.}(2008)\citenamefont {Krukau},
  \citenamefont {Scuseria}, \citenamefont {Perdew},\ and\ \citenamefont
  {Savin}}]{Krukau2008rsh}%
  \BibitemOpen
  \bibfield  {author} {\bibinfo {author} {\bibfnamefont {A.~V.}\ \bibnamefont
  {Krukau}}, \bibinfo {author} {\bibfnamefont {G.~E.}\ \bibnamefont
  {Scuseria}}, \bibinfo {author} {\bibfnamefont {J.~P.}\ \bibnamefont
  {Perdew}},\ and\ \bibinfo {author} {\bibfnamefont {A.}~\bibnamefont
  {Savin}},\ }\bibfield  {title} {\bibinfo {title} {Hybrid functionals with
  local range separation},\ }\href {https://doi.org/10.1063/1.2978377}
  {\bibfield  {journal} {\bibinfo  {journal} {The Journal of Chemical Physics}\
  }\textbf {\bibinfo {volume} {129}},\ \bibinfo {pages} {124103} (\bibinfo
  {year} {2008})}\BibitemShut {NoStop}%
\bibitem [{\citenamefont {Baer}\ and\ \citenamefont
  {Neuhauser}(2005)}]{baer_density_2005}%
  \BibitemOpen
  \bibfield  {author} {\bibinfo {author} {\bibfnamefont {R.}~\bibnamefont
  {Baer}}\ and\ \bibinfo {author} {\bibfnamefont {D.}~\bibnamefont
  {Neuhauser}},\ }\bibfield  {title} {\bibinfo {title} {Density functional
  theory with correct long-range asymptotic behavior},\ }\href
  {https://doi.org/10.1103/PhysRevLett.94.043002} {\bibfield  {journal}
  {\bibinfo  {journal} {Phys. Rev. Lett.}\ }\textbf {\bibinfo {volume} {94}},\
  \bibinfo {pages} {043002} (\bibinfo {year} {2005})}\BibitemShut {NoStop}%
\bibitem [{\citenamefont {Leininger}\ \emph {et~al.}(1997)\citenamefont
  {Leininger}, \citenamefont {Stoll}, \citenamefont {Werner},\ and\
  \citenamefont {Savin}}]{savin1997}%
  \BibitemOpen
  \bibfield  {author} {\bibinfo {author} {\bibfnamefont {T.}~\bibnamefont
  {Leininger}}, \bibinfo {author} {\bibfnamefont {H.}~\bibnamefont {Stoll}},
  \bibinfo {author} {\bibfnamefont {H.-J.}\ \bibnamefont {Werner}},\ and\
  \bibinfo {author} {\bibfnamefont {A.}~\bibnamefont {Savin}},\ }\bibfield
  {title} {\bibinfo {title} {Combining long-range configuration interaction
  with short-range density functionals},\ }\href
  {https://doi.org/https://doi.org/10.1016/S0009-2614(97)00758-6} {\bibfield
  {journal} {\bibinfo  {journal} {Chemical Physics Letters}\ }\textbf {\bibinfo
  {volume} {275}},\ \bibinfo {pages} {151} (\bibinfo {year}
  {1997})}\BibitemShut {NoStop}%
\bibitem [{\citenamefont {Ren}\ \emph {et~al.}(2012)\citenamefont {Ren},
  \citenamefont {Rinke}, \citenamefont {Blum}, \citenamefont {Wieferink},
  \citenamefont {Tkatchenko}, \citenamefont {Sanfilippo}, \citenamefont
  {Reuter},\ and\ \citenamefont {Scheffler}}]{ROI_2012}%
  \BibitemOpen
  \bibfield  {author} {\bibinfo {author} {\bibfnamefont {X.}~\bibnamefont
  {Ren}}, \bibinfo {author} {\bibfnamefont {P.}~\bibnamefont {Rinke}}, \bibinfo
  {author} {\bibfnamefont {V.}~\bibnamefont {Blum}}, \bibinfo {author}
  {\bibfnamefont {J.}~\bibnamefont {Wieferink}}, \bibinfo {author}
  {\bibfnamefont {A.}~\bibnamefont {Tkatchenko}}, \bibinfo {author}
  {\bibfnamefont {A.}~\bibnamefont {Sanfilippo}}, \bibinfo {author}
  {\bibfnamefont {K.}~\bibnamefont {Reuter}},\ and\ \bibinfo {author}
  {\bibfnamefont {M.}~\bibnamefont {Scheffler}},\ }\bibfield  {title} {\bibinfo
  {title} {Resolution-of-identity approach to {Hartree}–{Fock}, hybrid
  density functionals, {RPA}, {MP2} and {GW} with numeric atom-centered orbital
  basis functions},\ }\href {https://doi.org/10.1088/1367-2630/14/5/053020}
  {\bibfield  {journal} {\bibinfo  {journal} {New Journal of Physics}\ }\textbf
  {\bibinfo {volume} {14}},\ \bibinfo {pages} {053020} (\bibinfo {year}
  {2012})}\BibitemShut {NoStop}%
\bibitem [{\citenamefont {Mintmire}\ and\ \citenamefont
  {Dunlap}(1982)}]{Mintmire_1982}%
  \BibitemOpen
  \bibfield  {author} {\bibinfo {author} {\bibfnamefont {J.~W.}\ \bibnamefont
  {Mintmire}}\ and\ \bibinfo {author} {\bibfnamefont {B.~I.}\ \bibnamefont
  {Dunlap}},\ }\bibfield  {title} {\bibinfo {title} {Fitting the coulomb
  potential variationally in linear-combination-of-atomic-orbitals
  density-functional calculations},\ }\href
  {https://doi.org/10.1103/physreva.25.88} {\bibfield  {journal} {\bibinfo
  {journal} {Physical Review A}\ }\textbf {\bibinfo {volume} {25}},\ \bibinfo
  {pages} {88} (\bibinfo {year} {1982})}\BibitemShut {NoStop}%
\bibitem [{\citenamefont {Vahtras}\ \emph {et~al.}(1993)\citenamefont
  {Vahtras}, \citenamefont {Almlöf},\ and\ \citenamefont
  {Feyereisen}}]{Vahtras_1993}%
  \BibitemOpen
  \bibfield  {author} {\bibinfo {author} {\bibfnamefont {O.}~\bibnamefont
  {Vahtras}}, \bibinfo {author} {\bibfnamefont {J.}~\bibnamefont {Almlöf}},\
  and\ \bibinfo {author} {\bibfnamefont {M.}~\bibnamefont {Feyereisen}},\
  }\bibfield  {title} {\bibinfo {title} {Integral approximations for
  {LCAO}-{SCF} calculations},\ }\href
  {https://doi.org/10.1016/0009-2614(93)89151-7} {\bibfield  {journal}
  {\bibinfo  {journal} {Chemical Physics Letters}\ }\textbf {\bibinfo {volume}
  {213}},\ \bibinfo {pages} {514} (\bibinfo {year} {1993})}\BibitemShut
  {NoStop}%
\bibitem [{\citenamefont {Jung}\ \emph {et~al.}(2005)\citenamefont {Jung},
  \citenamefont {Sodt}, \citenamefont {Gill},\ and\ \citenamefont
  {Head-Gordon}}]{Jung_2005}%
  \BibitemOpen
  \bibfield  {author} {\bibinfo {author} {\bibfnamefont {Y.}~\bibnamefont
  {Jung}}, \bibinfo {author} {\bibfnamefont {A.}~\bibnamefont {Sodt}}, \bibinfo
  {author} {\bibfnamefont {P.~M.~W.}\ \bibnamefont {Gill}},\ and\ \bibinfo
  {author} {\bibfnamefont {M.}~\bibnamefont {Head-Gordon}},\ }\bibfield
  {title} {\bibinfo {title} {Auxiliary basis expansions for large-scale
  electronic structure calculations},\ }\href
  {https://doi.org/10.1073/pnas.0408475102} {\bibfield  {journal} {\bibinfo
  {journal} {Proceedings of the National Academy of Sciences}\ }\textbf
  {\bibinfo {volume} {102}},\ \bibinfo {pages} {6692} (\bibinfo {year}
  {2005})}\BibitemShut {NoStop}%
\bibitem [{\citenamefont {Damle}\ \emph {et~al.}(2015)\citenamefont {Damle},
  \citenamefont {Lin},\ and\ \citenamefont {Ying}}]{SCDM_2015}%
  \BibitemOpen
  \bibfield  {author} {\bibinfo {author} {\bibfnamefont {A.}~\bibnamefont
  {Damle}}, \bibinfo {author} {\bibfnamefont {L.}~\bibnamefont {Lin}},\ and\
  \bibinfo {author} {\bibfnamefont {L.}~\bibnamefont {Ying}},\ }\bibfield
  {title} {\bibinfo {title} {Compressed representation of {Kohn}–{Sham}
  orbitals via selected columns of the density matrix},\ }\href
  {https://doi.org/10.1021/ct500985f} {\bibfield  {journal} {\bibinfo
  {journal} {Journal of Chemical Theory and Computation}\ }\textbf {\bibinfo
  {volume} {11}},\ \bibinfo {pages} {1463} (\bibinfo {year}
  {2015})}\BibitemShut {NoStop}%
\bibitem [{\citenamefont {Giovannini}\ and\ \citenamefont
  {Koch}(2022)}]{tommaso_fragment2022}%
  \BibitemOpen
  \bibfield  {author} {\bibinfo {author} {\bibfnamefont {T.}~\bibnamefont
  {Giovannini}}\ and\ \bibinfo {author} {\bibfnamefont {H.}~\bibnamefont
  {Koch}},\ }\bibfield  {title} {\bibinfo {title} {Fragment localized molecular
  orbitals},\ }\href {https://doi.org/10.1021/acs.jctc.2c00359} {\bibfield
  {journal} {\bibinfo  {journal} {Journal of Chemical Theory and Computation}\
  }\textbf {\bibinfo {volume} {18}},\ \bibinfo {pages} {4806} (\bibinfo {year}
  {2022})}\BibitemShut {NoStop}%
\bibitem [{\citenamefont {Prentice}\ \emph {et~al.}(2020)\citenamefont
  {Prentice}, \citenamefont {Aarons}, \citenamefont {Womack}, \citenamefont
  {Allen}, \citenamefont {Andrinopoulos}, \citenamefont {Anton}, \citenamefont
  {Bell}, \citenamefont {Bhandari}, \citenamefont {Bramley}, \citenamefont
  {Charlton}, \citenamefont {Clements}, \citenamefont {Cole}, \citenamefont
  {Constantinescu}, \citenamefont {Corsetti}, \citenamefont {Dubois},
  \citenamefont {Duff}, \citenamefont {Escart{\'{\i}}n}, \citenamefont {Greco},
  \citenamefont {Hill}, \citenamefont {Lee}, \citenamefont {Linscott},
  \citenamefont {O'Regan}, \citenamefont {Phipps}, \citenamefont {Ratcliff},
  \citenamefont {Serrano}, \citenamefont {Tait}, \citenamefont {Teobaldi},
  \citenamefont {Vitale}, \citenamefont {Yeung}, \citenamefont {Zuehlsdorff},
  \citenamefont {Dziedzic}, \citenamefont {Haynes}, \citenamefont {Hine},
  \citenamefont {Mostofi}, \citenamefont {Payne},\ and\ \citenamefont
  {Skylaris}}]{Prentice_2020}%
  \BibitemOpen
  \bibfield  {author} {\bibinfo {author} {\bibfnamefont {J.~C.~A.}\
  \bibnamefont {Prentice}}, \bibinfo {author} {\bibfnamefont {J.}~\bibnamefont
  {Aarons}}, \bibinfo {author} {\bibfnamefont {J.~C.}\ \bibnamefont {Womack}},
  \bibinfo {author} {\bibfnamefont {A.~E.~A.}\ \bibnamefont {Allen}}, \bibinfo
  {author} {\bibfnamefont {L.}~\bibnamefont {Andrinopoulos}}, \bibinfo {author}
  {\bibfnamefont {L.}~\bibnamefont {Anton}}, \bibinfo {author} {\bibfnamefont
  {R.~A.}\ \bibnamefont {Bell}}, \bibinfo {author} {\bibfnamefont
  {A.}~\bibnamefont {Bhandari}}, \bibinfo {author} {\bibfnamefont {G.~A.}\
  \bibnamefont {Bramley}}, \bibinfo {author} {\bibfnamefont {R.~J.}\
  \bibnamefont {Charlton}}, \bibinfo {author} {\bibfnamefont {R.~J.}\
  \bibnamefont {Clements}}, \bibinfo {author} {\bibfnamefont {D.~J.}\
  \bibnamefont {Cole}}, \bibinfo {author} {\bibfnamefont {G.}~\bibnamefont
  {Constantinescu}}, \bibinfo {author} {\bibfnamefont {F.}~\bibnamefont
  {Corsetti}}, \bibinfo {author} {\bibfnamefont {S.~M.-M.}\ \bibnamefont
  {Dubois}}, \bibinfo {author} {\bibfnamefont {K.~K.~B.}\ \bibnamefont {Duff}},
  \bibinfo {author} {\bibfnamefont {J.~M.}\ \bibnamefont {Escart{\'{\i}}n}},
  \bibinfo {author} {\bibfnamefont {A.}~\bibnamefont {Greco}}, \bibinfo
  {author} {\bibfnamefont {Q.}~\bibnamefont {Hill}}, \bibinfo {author}
  {\bibfnamefont {L.~P.}\ \bibnamefont {Lee}}, \bibinfo {author} {\bibfnamefont
  {E.}~\bibnamefont {Linscott}}, \bibinfo {author} {\bibfnamefont {D.~D.}\
  \bibnamefont {O'Regan}}, \bibinfo {author} {\bibfnamefont {M.~J.~S.}\
  \bibnamefont {Phipps}}, \bibinfo {author} {\bibfnamefont {L.~E.}\
  \bibnamefont {Ratcliff}}, \bibinfo {author} {\bibfnamefont {{\'{A}}.~R.}\
  \bibnamefont {Serrano}}, \bibinfo {author} {\bibfnamefont {E.~W.}\
  \bibnamefont {Tait}}, \bibinfo {author} {\bibfnamefont {G.}~\bibnamefont
  {Teobaldi}}, \bibinfo {author} {\bibfnamefont {V.}~\bibnamefont {Vitale}},
  \bibinfo {author} {\bibfnamefont {N.}~\bibnamefont {Yeung}}, \bibinfo
  {author} {\bibfnamefont {T.~J.}\ \bibnamefont {Zuehlsdorff}}, \bibinfo
  {author} {\bibfnamefont {J.}~\bibnamefont {Dziedzic}}, \bibinfo {author}
  {\bibfnamefont {P.~D.}\ \bibnamefont {Haynes}}, \bibinfo {author}
  {\bibfnamefont {N.~D.~M.}\ \bibnamefont {Hine}}, \bibinfo {author}
  {\bibfnamefont {A.~A.}\ \bibnamefont {Mostofi}}, \bibinfo {author}
  {\bibfnamefont {M.~C.}\ \bibnamefont {Payne}},\ and\ \bibinfo {author}
  {\bibfnamefont {C.-K.}\ \bibnamefont {Skylaris}},\ }\bibfield  {title}
  {\bibinfo {title} {The {ONETEP} linear-scaling density functional theory
  program},\ }\bibfield  {journal} {\bibinfo  {journal} {The Journal of
  Chemical Physics}\ }\textbf {\bibinfo {volume} {152}},\ \href
  {https://doi.org/10.1063/5.0004445} {10.1063/5.0004445} (\bibinfo {year}
  {2020})\BibitemShut {NoStop}%
\bibitem [{\citenamefont {Graf}\ \emph {et~al.}(2018)\citenamefont {Graf},
  \citenamefont {Beuerle}, \citenamefont {Schurkus}, \citenamefont {Luenser},
  \citenamefont {Savasci},\ and\ \citenamefont {Ochsenfeld}}]{Graf_2018}%
  \BibitemOpen
  \bibfield  {author} {\bibinfo {author} {\bibfnamefont {D.}~\bibnamefont
  {Graf}}, \bibinfo {author} {\bibfnamefont {M.}~\bibnamefont {Beuerle}},
  \bibinfo {author} {\bibfnamefont {H.~F.}\ \bibnamefont {Schurkus}}, \bibinfo
  {author} {\bibfnamefont {A.}~\bibnamefont {Luenser}}, \bibinfo {author}
  {\bibfnamefont {G.}~\bibnamefont {Savasci}},\ and\ \bibinfo {author}
  {\bibfnamefont {C.}~\bibnamefont {Ochsenfeld}},\ }\bibfield  {title}
  {\bibinfo {title} {Accurate and efficient parallel implementation of an
  effective linear-scaling direct random phase approximation method},\ }\href
  {https://doi.org/10.1021/acs.jctc.8b00177} {\bibfield  {journal} {\bibinfo
  {journal} {Journal of Chemical Theory and Computation}\ }\textbf {\bibinfo
  {volume} {14}},\ \bibinfo {pages} {2505} (\bibinfo {year}
  {2018})}\BibitemShut {NoStop}%
\bibitem [{\citenamefont {Ufimtsev}\ and\ \citenamefont
  {Mart{\'{\i}}nez}(2008)}]{Martinez2008GPU}%
  \BibitemOpen
  \bibfield  {author} {\bibinfo {author} {\bibfnamefont {I.~S.}\ \bibnamefont
  {Ufimtsev}}\ and\ \bibinfo {author} {\bibfnamefont {T.~J.}\ \bibnamefont
  {Mart{\'{\i}}nez}},\ }\bibfield  {title} {\bibinfo {title} {Quantum chemistry
  on graphical processing units. 1. {Strategies} for {Two}-{Electron}
  {Integral} {Evaluation}},\ }\href {https://doi.org/10.1021/ct700268q}
  {\bibfield  {journal} {\bibinfo  {journal} {Journal of Chemical Theory and
  Computation}\ }\textbf {\bibinfo {volume} {4}},\ \bibinfo {pages} {222}
  (\bibinfo {year} {2008})}\BibitemShut {NoStop}%
\bibitem [{\citenamefont {Kussmann}\ \emph {et~al.}(2021)\citenamefont
  {Kussmann}, \citenamefont {Laqua},\ and\ \citenamefont
  {Ochsenfeld}}]{Kussmann_2021}%
  \BibitemOpen
  \bibfield  {author} {\bibinfo {author} {\bibfnamefont {J.}~\bibnamefont
  {Kussmann}}, \bibinfo {author} {\bibfnamefont {H.}~\bibnamefont {Laqua}},\
  and\ \bibinfo {author} {\bibfnamefont {C.}~\bibnamefont {Ochsenfeld}},\
  }\bibfield  {title} {\bibinfo {title} {Highly efficient
  resolution-of-identity density functional theory calculations on central and
  graphics processing units},\ }\href
  {https://doi.org/10.1021/acs.jctc.0c01252} {\bibfield  {journal} {\bibinfo
  {journal} {Journal of Chemical Theory and Computation}\ }\textbf {\bibinfo
  {volume} {17}},\ \bibinfo {pages} {1512} (\bibinfo {year}
  {2021})}\BibitemShut {NoStop}%
\bibitem [{\citenamefont {Neuhauser}\ \emph {et~al.}(2016)\citenamefont
  {Neuhauser}, \citenamefont {Rabani}, \citenamefont {Cytter},\ and\
  \citenamefont {Baer}}]{neuhauser2016stochastic_exchange}%
  \BibitemOpen
  \bibfield  {author} {\bibinfo {author} {\bibfnamefont {D.}~\bibnamefont
  {Neuhauser}}, \bibinfo {author} {\bibfnamefont {E.}~\bibnamefont {Rabani}},
  \bibinfo {author} {\bibfnamefont {Y.}~\bibnamefont {Cytter}},\ and\ \bibinfo
  {author} {\bibfnamefont {R.}~\bibnamefont {Baer}},\ }\bibfield  {title}
  {\bibinfo {title} {Stochastic optimally tuned range-separated hybrid density
  functional theory},\ }\href {https://doi.org/10.1021/acs.jpca.5b10573}
  {\bibfield  {journal} {\bibinfo  {journal} {The Journal of Physical Chemistry
  A}\ }\textbf {\bibinfo {volume} {120}},\ \bibinfo {pages} {3071} (\bibinfo
  {year} {2016})}\BibitemShut {NoStop}%
\bibitem [{\citenamefont {Vlček}\ \emph {et~al.}(2018)\citenamefont {Vlček},
  \citenamefont {Li}, \citenamefont {Baer}, \citenamefont {Rabani},\ and\
  \citenamefont {Neuhauser}}]{vlvcek2018swiftg}%
  \BibitemOpen
  \bibfield  {author} {\bibinfo {author} {\bibfnamefont {V.}~\bibnamefont
  {Vlček}}, \bibinfo {author} {\bibfnamefont {W.}~\bibnamefont {Li}}, \bibinfo
  {author} {\bibfnamefont {R.}~\bibnamefont {Baer}}, \bibinfo {author}
  {\bibfnamefont {E.}~\bibnamefont {Rabani}},\ and\ \bibinfo {author}
  {\bibfnamefont {D.}~\bibnamefont {Neuhauser}},\ }\bibfield  {title} {\bibinfo
  {title} {Swift ${GW}$ beyond 10,000 electrons using sparse stochastic
  compression},\ }\href {https://doi.org/10.1103/PhysRevB.98.075107} {\bibfield
   {journal} {\bibinfo  {journal} {Phys. Rev. B}\ }\textbf {\bibinfo {volume}
  {98}},\ \bibinfo {pages} {075107} (\bibinfo {year} {2018})}\BibitemShut
  {NoStop}%
\bibitem [{\citenamefont {Dou}\ \emph {et~al.}(2020)\citenamefont {Dou},
  \citenamefont {Chen}, \citenamefont {Takeshita}, \citenamefont {Baer},
  \citenamefont {Neuhauser},\ and\ \citenamefont {Rabani}}]{SR-SRI_2020}%
  \BibitemOpen
  \bibfield  {author} {\bibinfo {author} {\bibfnamefont {W.}~\bibnamefont
  {Dou}}, \bibinfo {author} {\bibfnamefont {M.}~\bibnamefont {Chen}}, \bibinfo
  {author} {\bibfnamefont {T.~Y.}\ \bibnamefont {Takeshita}}, \bibinfo {author}
  {\bibfnamefont {R.}~\bibnamefont {Baer}}, \bibinfo {author} {\bibfnamefont
  {D.}~\bibnamefont {Neuhauser}},\ and\ \bibinfo {author} {\bibfnamefont
  {E.}~\bibnamefont {Rabani}},\ }\bibfield  {title} {\bibinfo {title}
  {{Range-separated stochastic resolution of identity: Formulation and
  application to second-order Green’s function theory}},\ }\href
  {https://doi.org/10.1063/5.0015177} {\bibfield  {journal} {\bibinfo
  {journal} {The Journal of Chemical Physics}\ }\textbf {\bibinfo {volume}
  {153}},\ \bibinfo {pages} {074113} (\bibinfo {year} {2020})}\BibitemShut
  {NoStop}%
\bibitem [{\citenamefont {Perdew}\ and\ \citenamefont
  {Wang}(1992)}]{PerdewWang1992}%
  \BibitemOpen
  \bibfield  {author} {\bibinfo {author} {\bibfnamefont {J.~P.}\ \bibnamefont
  {Perdew}}\ and\ \bibinfo {author} {\bibfnamefont {Y.}~\bibnamefont {Wang}},\
  }\bibfield  {title} {\bibinfo {title} {Accurate and simple analytic
  representation of the electron-gas correlation energy},\ }\href
  {https://doi.org/10.1103/PhysRevB.45.13244} {\bibfield  {journal} {\bibinfo
  {journal} {Phys. Rev. B}\ }\textbf {\bibinfo {volume} {45}},\ \bibinfo
  {pages} {13244} (\bibinfo {year} {1992})}\BibitemShut {NoStop}%
\bibitem [{\citenamefont {Troullier}\ and\ \citenamefont
  {Martins}(1991)}]{TroullierMartins91}%
  \BibitemOpen
  \bibfield  {author} {\bibinfo {author} {\bibfnamefont {N.}~\bibnamefont
  {Troullier}}\ and\ \bibinfo {author} {\bibfnamefont {J.~L.}\ \bibnamefont
  {Martins}},\ }\bibfield  {title} {\bibinfo {title} {Efficient
  pseudopotentials for plane-wave calculations},\ }\href
  {https://doi.org/10.1103/PhysRevB.43.1993} {\bibfield  {journal} {\bibinfo
  {journal} {Phys. Rev. B}\ }\textbf {\bibinfo {volume} {43}},\ \bibinfo
  {pages} {1993} (\bibinfo {year} {1991})}\BibitemShut {NoStop}%
\bibitem [{\citenamefont {Martyna}\ and\ \citenamefont
  {Tuckerman}(1999)}]{MartynaTuckerman1999}%
  \BibitemOpen
  \bibfield  {author} {\bibinfo {author} {\bibfnamefont {G.~J.}\ \bibnamefont
  {Martyna}}\ and\ \bibinfo {author} {\bibfnamefont {M.~E.}\ \bibnamefont
  {Tuckerman}},\ }\bibfield  {title} {\bibinfo {title} {{A reciprocal space
  based method for treating long range interactions in ab initio and
  force-field-based calculations in clusters}},\ }\href
  {https://doi.org/10.1063/1.477923} {\bibfield  {journal} {\bibinfo  {journal}
  {The Journal of Chemical Physics}\ }\textbf {\bibinfo {volume} {110}},\
  \bibinfo {pages} {2810} (\bibinfo {year} {1999})}\BibitemShut {NoStop}%
\bibitem [{\citenamefont {Baer}\ \emph {et~al.}(2010)\citenamefont {Baer},
  \citenamefont {Livshits},\ and\ \citenamefont {Salzner}}]{baer_tuned_2010}%
  \BibitemOpen
  \bibfield  {author} {\bibinfo {author} {\bibfnamefont {R.}~\bibnamefont
  {Baer}}, \bibinfo {author} {\bibfnamefont {E.}~\bibnamefont {Livshits}},\
  and\ \bibinfo {author} {\bibfnamefont {U.}~\bibnamefont {Salzner}},\
  }\bibfield  {title} {\bibinfo {title} {Tuned range-separated hybrids in
  density functional theory},\ }\href
  {https://doi.org/10.1146/annurev.physchem.012809.103321} {\bibfield
  {journal} {\bibinfo  {journal} {Annual Review of Physical Chemistry}\
  }\textbf {\bibinfo {volume} {61}},\ \bibinfo {pages} {85} (\bibinfo {year}
  {2010})}\BibitemShut {NoStop}%
\bibitem [{\citenamefont {Förster}\ and\ \citenamefont
  {Visscher}(2022)}]{Forster_ADF_Chromo_2022}%
  \BibitemOpen
  \bibfield  {author} {\bibinfo {author} {\bibfnamefont {A.}~\bibnamefont
  {Förster}}\ and\ \bibinfo {author} {\bibfnamefont {L.}~\bibnamefont
  {Visscher}},\ }\bibfield  {title} {\bibinfo {title} {Quasiparticle
  self-consistent {GW}-{Bethe-Salpeter} equation calculations for large
  chromophoric systems},\ }\href {https://doi.org/10.1021/} {\bibfield
  {journal} {\bibinfo  {journal} {Journal of Chemical Theory and Computation}\
  }\textbf {\bibinfo {volume} {18}},\ \bibinfo {pages} {6779} (\bibinfo {year}
  {2022})}\BibitemShut {NoStop}%
\bibitem [{\citenamefont {Sirohiwal}\ and\ \citenamefont
  {Pantazis}(2022)}]{psn_cnts2022}%
  \BibitemOpen
  \bibfield  {author} {\bibinfo {author} {\bibfnamefont {A.}~\bibnamefont
  {Sirohiwal}}\ and\ \bibinfo {author} {\bibfnamefont {D.~A.}\ \bibnamefont
  {Pantazis}},\ }\bibfield  {title} {\bibinfo {title} {The electronic origin of
  far-red-light-driven oxygenic photosynthesis},\ }\href
  {https://doi.org/https://doi.org/10.1002/anie.202200356} {\bibfield
  {journal} {\bibinfo  {journal} {Angewandte Chemie International Edition}\
  }\textbf {\bibinfo {volume} {61}},\ \bibinfo {pages} {e202200356} (\bibinfo
  {year} {2022})}\BibitemShut {NoStop}%
\bibitem [{\citenamefont {Apr{\`a}}\ \emph {et~al.}(2020)\citenamefont
  {Apr{\`a}}, \citenamefont {Bylaska}, \citenamefont {de~Jong}, \citenamefont
  {Govind}, \citenamefont {Kowalski}, \citenamefont {Straatsma}, \citenamefont
  {Valiev}, \citenamefont {van Dam}, \citenamefont {Alexeev}, \citenamefont
  {Anchell}, \citenamefont {Anisimov}, \citenamefont {Aquino}, \citenamefont
  {Atta-Fynn}, \citenamefont {Autschbach}, \citenamefont {Bauman},
  \citenamefont {Becca}, \citenamefont {Bernholdt}, \citenamefont
  {Bhaskaran-Nair}, \citenamefont {Bogatko}, \citenamefont {Borowski},
  \citenamefont {Boschen}, \citenamefont {Brabec}, \citenamefont {Bruner},
  \citenamefont {Cau{\"e}t}, \citenamefont {Chen}, \citenamefont {Chuev},
  \citenamefont {Cramer}, \citenamefont {Daily}, \citenamefont {Deegan},
  \citenamefont {Dunning}, \citenamefont {Dupuis}, \citenamefont {Dyall},
  \citenamefont {Fann}, \citenamefont {Fischer}, \citenamefont {Fonari},
  \citenamefont {Fr{\"u}chtl}, \citenamefont {Gagliardi}, \citenamefont
  {Garza}, \citenamefont {Gawande}, \citenamefont {Ghosh}, \citenamefont
  {Glaesemann}, \citenamefont {G{\"o}tz}, \citenamefont {Hammond},
  \citenamefont {Helms}, \citenamefont {Hermes}, \citenamefont {Hirao},
  \citenamefont {Hirata}, \citenamefont {Jacquelin}, \citenamefont {Jensen},
  \citenamefont {Johnson}, \citenamefont {J{\'o}nsson}, \citenamefont
  {Kendall}, \citenamefont {Klemm}, \citenamefont {Kobayashi}, \citenamefont
  {Konkov}, \citenamefont {Krishnamoorthy}, \citenamefont {Krishnan},
  \citenamefont {Lin}, \citenamefont {Lins}, \citenamefont {Littlefield},
  \citenamefont {Logsdail}, \citenamefont {Lopata}, \citenamefont {Ma},
  \citenamefont {Marenich}, \citenamefont {Martin~del Campo}, \citenamefont
  {Mejia-Rodriguez}, \citenamefont {Moore}, \citenamefont {Mullin},
  \citenamefont {Nakajima}, \citenamefont {Nascimento}, \citenamefont
  {Nichols}, \citenamefont {Nichols}, \citenamefont {Nieplocha}, \citenamefont
  {Otero-de-la Roza}, \citenamefont {Palmer}, \citenamefont {Panyala},
  \citenamefont {Pirojsirikul}, \citenamefont {Peng}, \citenamefont {Peverati},
  \citenamefont {Pittner}, \citenamefont {Pollack}, \citenamefont {Richard},
  \citenamefont {Sadayappan}, \citenamefont {Schatz}, \citenamefont {Shelton},
  \citenamefont {Silverstein}, \citenamefont {Smith}, \citenamefont {Soares},
  \citenamefont {Song}, \citenamefont {Swart}, \citenamefont {Taylor},
  \citenamefont {Thomas}, \citenamefont {Tipparaju}, \citenamefont {Truhlar},
  \citenamefont {Tsemekhman}, \citenamefont {Van~Voorhis}, \citenamefont
  {V{\'a}zquez-Mayagoitia}, \citenamefont {Verma}, \citenamefont {Villa},
  \citenamefont {Vishnu}, \citenamefont {Vogiatzis}, \citenamefont {Wang},
  \citenamefont {Weare}, \citenamefont {Williamson}, \citenamefont {Windus},
  \citenamefont {Woli{\'n}ski}, \citenamefont {Wong}, \citenamefont {Wu},
  \citenamefont {Yang}, \citenamefont {Yu}, \citenamefont {Zacharias},
  \citenamefont {Zhang}, \citenamefont {Zhao},\ and\ \citenamefont
  {Harrison}}]{NWChem2020}%
  \BibitemOpen
  \bibfield  {author} {\bibinfo {author} {\bibfnamefont {E.}~\bibnamefont
  {Apr{\`a}}}, \bibinfo {author} {\bibfnamefont {E.~J.}\ \bibnamefont
  {Bylaska}}, \bibinfo {author} {\bibfnamefont {W.~A.}\ \bibnamefont
  {de~Jong}}, \bibinfo {author} {\bibfnamefont {N.}~\bibnamefont {Govind}},
  \bibinfo {author} {\bibfnamefont {K.}~\bibnamefont {Kowalski}}, \bibinfo
  {author} {\bibfnamefont {T.~P.}\ \bibnamefont {Straatsma}}, \bibinfo {author}
  {\bibfnamefont {M.}~\bibnamefont {Valiev}}, \bibinfo {author} {\bibfnamefont
  {H.~J.~J.}\ \bibnamefont {van Dam}}, \bibinfo {author} {\bibfnamefont
  {Y.}~\bibnamefont {Alexeev}}, \bibinfo {author} {\bibfnamefont
  {J.}~\bibnamefont {Anchell}}, \bibinfo {author} {\bibfnamefont
  {V.}~\bibnamefont {Anisimov}}, \bibinfo {author} {\bibfnamefont {F.~W.}\
  \bibnamefont {Aquino}}, \bibinfo {author} {\bibfnamefont {R.}~\bibnamefont
  {Atta-Fynn}}, \bibinfo {author} {\bibfnamefont {J.}~\bibnamefont
  {Autschbach}}, \bibinfo {author} {\bibfnamefont {N.~P.}\ \bibnamefont
  {Bauman}}, \bibinfo {author} {\bibfnamefont {J.~C.}\ \bibnamefont {Becca}},
  \bibinfo {author} {\bibfnamefont {D.~E.}\ \bibnamefont {Bernholdt}}, \bibinfo
  {author} {\bibfnamefont {K.}~\bibnamefont {Bhaskaran-Nair}}, \bibinfo
  {author} {\bibfnamefont {S.}~\bibnamefont {Bogatko}}, \bibinfo {author}
  {\bibfnamefont {P.}~\bibnamefont {Borowski}}, \bibinfo {author}
  {\bibfnamefont {J.}~\bibnamefont {Boschen}}, \bibinfo {author} {\bibfnamefont
  {J.}~\bibnamefont {Brabec}}, \bibinfo {author} {\bibfnamefont
  {A.}~\bibnamefont {Bruner}}, \bibinfo {author} {\bibfnamefont
  {E.}~\bibnamefont {Cau{\"e}t}}, \bibinfo {author} {\bibfnamefont
  {Y.}~\bibnamefont {Chen}}, \bibinfo {author} {\bibfnamefont {G.~N.}\
  \bibnamefont {Chuev}}, \bibinfo {author} {\bibfnamefont {C.~J.}\ \bibnamefont
  {Cramer}}, \bibinfo {author} {\bibfnamefont {J.}~\bibnamefont {Daily}},
  \bibinfo {author} {\bibfnamefont {M.~J.~O.}\ \bibnamefont {Deegan}}, \bibinfo
  {author} {\bibfnamefont {J.}~\bibnamefont {Dunning}, \bibfnamefont {T.~H.}},
  \bibinfo {author} {\bibfnamefont {M.}~\bibnamefont {Dupuis}}, \bibinfo
  {author} {\bibfnamefont {K.~G.}\ \bibnamefont {Dyall}}, \bibinfo {author}
  {\bibfnamefont {G.~I.}\ \bibnamefont {Fann}}, \bibinfo {author}
  {\bibfnamefont {S.~A.}\ \bibnamefont {Fischer}}, \bibinfo {author}
  {\bibfnamefont {A.}~\bibnamefont {Fonari}}, \bibinfo {author} {\bibfnamefont
  {H.}~\bibnamefont {Fr{\"u}chtl}}, \bibinfo {author} {\bibfnamefont
  {L.}~\bibnamefont {Gagliardi}}, \bibinfo {author} {\bibfnamefont
  {J.}~\bibnamefont {Garza}}, \bibinfo {author} {\bibfnamefont
  {N.}~\bibnamefont {Gawande}}, \bibinfo {author} {\bibfnamefont
  {S.}~\bibnamefont {Ghosh}}, \bibinfo {author} {\bibfnamefont
  {K.}~\bibnamefont {Glaesemann}}, \bibinfo {author} {\bibfnamefont {A.~W.}\
  \bibnamefont {G{\"o}tz}}, \bibinfo {author} {\bibfnamefont {J.}~\bibnamefont
  {Hammond}}, \bibinfo {author} {\bibfnamefont {V.}~\bibnamefont {Helms}},
  \bibinfo {author} {\bibfnamefont {E.~D.}\ \bibnamefont {Hermes}}, \bibinfo
  {author} {\bibfnamefont {K.}~\bibnamefont {Hirao}}, \bibinfo {author}
  {\bibfnamefont {S.}~\bibnamefont {Hirata}}, \bibinfo {author} {\bibfnamefont
  {M.}~\bibnamefont {Jacquelin}}, \bibinfo {author} {\bibfnamefont
  {L.}~\bibnamefont {Jensen}}, \bibinfo {author} {\bibfnamefont {B.~G.}\
  \bibnamefont {Johnson}}, \bibinfo {author} {\bibfnamefont {H.}~\bibnamefont
  {J{\'o}nsson}}, \bibinfo {author} {\bibfnamefont {R.~A.}\ \bibnamefont
  {Kendall}}, \bibinfo {author} {\bibfnamefont {M.}~\bibnamefont {Klemm}},
  \bibinfo {author} {\bibfnamefont {R.}~\bibnamefont {Kobayashi}}, \bibinfo
  {author} {\bibfnamefont {V.}~\bibnamefont {Konkov}}, \bibinfo {author}
  {\bibfnamefont {S.}~\bibnamefont {Krishnamoorthy}}, \bibinfo {author}
  {\bibfnamefont {M.}~\bibnamefont {Krishnan}}, \bibinfo {author}
  {\bibfnamefont {Z.}~\bibnamefont {Lin}}, \bibinfo {author} {\bibfnamefont
  {R.~D.}\ \bibnamefont {Lins}}, \bibinfo {author} {\bibfnamefont {R.~J.}\
  \bibnamefont {Littlefield}}, \bibinfo {author} {\bibfnamefont {A.~J.}\
  \bibnamefont {Logsdail}}, \bibinfo {author} {\bibfnamefont {K.}~\bibnamefont
  {Lopata}}, \bibinfo {author} {\bibfnamefont {W.}~\bibnamefont {Ma}}, \bibinfo
  {author} {\bibfnamefont {A.~V.}\ \bibnamefont {Marenich}}, \bibinfo {author}
  {\bibfnamefont {J.}~\bibnamefont {Martin~del Campo}}, \bibinfo {author}
  {\bibfnamefont {D.}~\bibnamefont {Mejia-Rodriguez}}, \bibinfo {author}
  {\bibfnamefont {J.~E.}\ \bibnamefont {Moore}}, \bibinfo {author}
  {\bibfnamefont {J.~M.}\ \bibnamefont {Mullin}}, \bibinfo {author}
  {\bibfnamefont {T.}~\bibnamefont {Nakajima}}, \bibinfo {author}
  {\bibfnamefont {D.~R.}\ \bibnamefont {Nascimento}}, \bibinfo {author}
  {\bibfnamefont {J.~A.}\ \bibnamefont {Nichols}}, \bibinfo {author}
  {\bibfnamefont {P.~J.}\ \bibnamefont {Nichols}}, \bibinfo {author}
  {\bibfnamefont {J.}~\bibnamefont {Nieplocha}}, \bibinfo {author}
  {\bibfnamefont {A.}~\bibnamefont {Otero-de-la Roza}}, \bibinfo {author}
  {\bibfnamefont {B.}~\bibnamefont {Palmer}}, \bibinfo {author} {\bibfnamefont
  {A.}~\bibnamefont {Panyala}}, \bibinfo {author} {\bibfnamefont
  {T.}~\bibnamefont {Pirojsirikul}}, \bibinfo {author} {\bibfnamefont
  {B.}~\bibnamefont {Peng}}, \bibinfo {author} {\bibfnamefont {R.}~\bibnamefont
  {Peverati}}, \bibinfo {author} {\bibfnamefont {J.}~\bibnamefont {Pittner}},
  \bibinfo {author} {\bibfnamefont {L.}~\bibnamefont {Pollack}}, \bibinfo
  {author} {\bibfnamefont {R.~M.}\ \bibnamefont {Richard}}, \bibinfo {author}
  {\bibfnamefont {P.}~\bibnamefont {Sadayappan}}, \bibinfo {author}
  {\bibfnamefont {G.~C.}\ \bibnamefont {Schatz}}, \bibinfo {author}
  {\bibfnamefont {W.~A.}\ \bibnamefont {Shelton}}, \bibinfo {author}
  {\bibfnamefont {D.~W.}\ \bibnamefont {Silverstein}}, \bibinfo {author}
  {\bibfnamefont {D.~M.~A.}\ \bibnamefont {Smith}}, \bibinfo {author}
  {\bibfnamefont {T.~A.}\ \bibnamefont {Soares}}, \bibinfo {author}
  {\bibfnamefont {D.}~\bibnamefont {Song}}, \bibinfo {author} {\bibfnamefont
  {M.}~\bibnamefont {Swart}}, \bibinfo {author} {\bibfnamefont {H.~L.}\
  \bibnamefont {Taylor}}, \bibinfo {author} {\bibfnamefont {G.~S.}\
  \bibnamefont {Thomas}}, \bibinfo {author} {\bibfnamefont {V.}~\bibnamefont
  {Tipparaju}}, \bibinfo {author} {\bibfnamefont {D.~G.}\ \bibnamefont
  {Truhlar}}, \bibinfo {author} {\bibfnamefont {K.}~\bibnamefont {Tsemekhman}},
  \bibinfo {author} {\bibfnamefont {T.}~\bibnamefont {Van~Voorhis}}, \bibinfo
  {author} {\bibfnamefont {{\'A}.}~\bibnamefont {V{\'a}zquez-Mayagoitia}},
  \bibinfo {author} {\bibfnamefont {P.}~\bibnamefont {Verma}}, \bibinfo
  {author} {\bibfnamefont {O.}~\bibnamefont {Villa}}, \bibinfo {author}
  {\bibfnamefont {A.}~\bibnamefont {Vishnu}}, \bibinfo {author} {\bibfnamefont
  {K.~D.}\ \bibnamefont {Vogiatzis}}, \bibinfo {author} {\bibfnamefont
  {D.}~\bibnamefont {Wang}}, \bibinfo {author} {\bibfnamefont {J.~H.}\
  \bibnamefont {Weare}}, \bibinfo {author} {\bibfnamefont {M.~J.}\ \bibnamefont
  {Williamson}}, \bibinfo {author} {\bibfnamefont {T.~L.}\ \bibnamefont
  {Windus}}, \bibinfo {author} {\bibfnamefont {K.}~\bibnamefont
  {Woli{\'n}ski}}, \bibinfo {author} {\bibfnamefont {A.~T.}\ \bibnamefont
  {Wong}}, \bibinfo {author} {\bibfnamefont {Q.}~\bibnamefont {Wu}}, \bibinfo
  {author} {\bibfnamefont {C.}~\bibnamefont {Yang}}, \bibinfo {author}
  {\bibfnamefont {Q.}~\bibnamefont {Yu}}, \bibinfo {author} {\bibfnamefont
  {M.}~\bibnamefont {Zacharias}}, \bibinfo {author} {\bibfnamefont
  {Z.}~\bibnamefont {Zhang}}, \bibinfo {author} {\bibfnamefont
  {Y.}~\bibnamefont {Zhao}},\ and\ \bibinfo {author} {\bibfnamefont {R.~J.}\
  \bibnamefont {Harrison}},\ }\bibfield  {title} {\bibinfo {title} {{NWChem:
  Past, present, and future}},\ }\href {https://doi.org/10.1063/5.0004997}
  {\bibfield  {journal} {\bibinfo  {journal} {The Journal of Chemical Physics}\
  }\textbf {\bibinfo {volume} {152}},\ \bibinfo {pages} {184102} (\bibinfo
  {year} {2020})}\BibitemShut {NoStop}%
\bibitem [{\citenamefont {Vlcek}\ \emph {et~al.}(2018)\citenamefont {Vlcek},
  \citenamefont {Baer}, \citenamefont {Rabani},\ and\ \citenamefont
  {Neuhauser}}]{Vlek2018scissors}%
  \BibitemOpen
  \bibfield  {author} {\bibinfo {author} {\bibfnamefont {V.}~\bibnamefont
  {Vlcek}}, \bibinfo {author} {\bibfnamefont {R.}~\bibnamefont {Baer}},
  \bibinfo {author} {\bibfnamefont {E.}~\bibnamefont {Rabani}},\ and\ \bibinfo
  {author} {\bibfnamefont {D.}~\bibnamefont {Neuhauser}},\ }\bibfield  {title}
  {\bibinfo {title} {Simple eigenvalue-self-consistent {$\Delta GW_0$}},\
  }\href {https://doi.org/10.1063/1.5042785} {\bibfield  {journal} {\bibinfo
  {journal} {J. Chem. Phys.}\ }\textbf {\bibinfo {volume} {149}},\ \bibinfo
  {pages} {174107} (\bibinfo {year} {2018})}\BibitemShut {NoStop}%
\bibitem [{\citenamefont {Neuhauser}\ \emph
  {et~al.}(2014{\natexlab{a}})\citenamefont {Neuhauser}, \citenamefont {Gao},
  \citenamefont {Arntsen}, \citenamefont {Karshenas}, \citenamefont {Rabani},\
  and\ \citenamefont {Baer}}]{neuhauser2014breaking}%
  \BibitemOpen
  \bibfield  {author} {\bibinfo {author} {\bibfnamefont {D.}~\bibnamefont
  {Neuhauser}}, \bibinfo {author} {\bibfnamefont {Y.}~\bibnamefont {Gao}},
  \bibinfo {author} {\bibfnamefont {C.}~\bibnamefont {Arntsen}}, \bibinfo
  {author} {\bibfnamefont {C.}~\bibnamefont {Karshenas}}, \bibinfo {author}
  {\bibfnamefont {E.}~\bibnamefont {Rabani}},\ and\ \bibinfo {author}
  {\bibfnamefont {R.}~\bibnamefont {Baer}},\ }\bibfield  {title} {\bibinfo
  {title} {Breaking the theoretical scaling limit for predicting quasiparticle
  energies: The stochastic {GW} approach},\ }\href@noop {} {\bibfield
  {journal} {\bibinfo  {journal} {Phys. Rev. Lett.}\ }\textbf {\bibinfo
  {volume} {113}},\ \bibinfo {pages} {076402} (\bibinfo {year}
  {2014}{\natexlab{a}})}\BibitemShut {NoStop}%
\bibitem [{\citenamefont {Vlcek}\ \emph {et~al.}(2019)\citenamefont {Vlcek},
  \citenamefont {Rabani}, \citenamefont {Baer},\ and\ \citenamefont
  {Neuhauser}}]{vlvcek2019nonmonotonic}%
  \BibitemOpen
  \bibfield  {author} {\bibinfo {author} {\bibfnamefont {V.}~\bibnamefont
  {Vlcek}}, \bibinfo {author} {\bibfnamefont {E.}~\bibnamefont {Rabani}},
  \bibinfo {author} {\bibfnamefont {R.}~\bibnamefont {Baer}},\ and\ \bibinfo
  {author} {\bibfnamefont {D.}~\bibnamefont {Neuhauser}},\ }\bibfield  {title}
  {\bibinfo {title} {Nonmonotonic band gap evolution in bent phosphorene
  nanosheets},\ }\href {https://doi.org/10.1103/PhysRevMaterials.3.064601}
  {\bibfield  {journal} {\bibinfo  {journal} {Phys. Rev. Mater.}\ }\textbf
  {\bibinfo {volume} {3}},\ \bibinfo {pages} {064601} (\bibinfo {year}
  {2019})}\BibitemShut {NoStop}%
\bibitem [{\citenamefont {Bruneval}\ and\ \citenamefont
  {Marques}(2012)}]{Bruneval_2012}%
  \BibitemOpen
  \bibfield  {author} {\bibinfo {author} {\bibfnamefont {F.}~\bibnamefont
  {Bruneval}}\ and\ \bibinfo {author} {\bibfnamefont {M.~A.~L.}\ \bibnamefont
  {Marques}},\ }\bibfield  {title} {\bibinfo {title} {Benchmarking the starting
  points of the {GW} approximation for molecules},\ }\href
  {https://doi.org/10.1021/ct300835h} {\bibfield  {journal} {\bibinfo
  {journal} {Journal of Chemical Theory and Computation}\ }\textbf {\bibinfo
  {volume} {9}},\ \bibinfo {pages} {324} (\bibinfo {year} {2012})}\BibitemShut
  {NoStop}%
\bibitem [{\citenamefont {McKeon}\ \emph {et~al.}(2022)\citenamefont {McKeon},
  \citenamefont {Hamed}, \citenamefont {Bruneval},\ and\ \citenamefont
  {Neaton}}]{McKeon2022}%
  \BibitemOpen
  \bibfield  {author} {\bibinfo {author} {\bibfnamefont {C.~A.}\ \bibnamefont
  {McKeon}}, \bibinfo {author} {\bibfnamefont {S.~M.}\ \bibnamefont {Hamed}},
  \bibinfo {author} {\bibfnamefont {F.}~\bibnamefont {Bruneval}},\ and\
  \bibinfo {author} {\bibfnamefont {J.~B.}\ \bibnamefont {Neaton}},\ }\bibfield
   {title} {\bibinfo {title} {An optimally tuned range-separated hybrid
  starting point for {GW} plus {Bethe}{\textendash}{Salpeter} equation
  calculations of molecules},\ }\bibfield  {journal} {\bibinfo  {journal} {The
  Journal of Chemical Physics}\ }\textbf {\bibinfo {volume} {157}},\ \href
  {https://doi.org/10.1063/5.0097582} {10.1063/5.0097582} (\bibinfo {year}
  {2022})\BibitemShut {NoStop}%
\bibitem [{\citenamefont {Bradbury}\ \emph {et~al.}(2022)\citenamefont
  {Bradbury}, \citenamefont {Nguyen}, \citenamefont {Caram},\ and\
  \citenamefont {Neuhauser}}]{bradbury_bse_2022}%
  \BibitemOpen
  \bibfield  {author} {\bibinfo {author} {\bibfnamefont {N.~C.}\ \bibnamefont
  {Bradbury}}, \bibinfo {author} {\bibfnamefont {M.}~\bibnamefont {Nguyen}},
  \bibinfo {author} {\bibfnamefont {J.~R.}\ \bibnamefont {Caram}},\ and\
  \bibinfo {author} {\bibfnamefont {D.}~\bibnamefont {Neuhauser}},\ }\bibfield
  {title} {\bibinfo {title} {{Bethe–Salpeter equation spectra for very large
  systems}},\ }\href {https://doi.org/10.1063/5.0100213} {\bibfield  {journal}
  {\bibinfo  {journal} {The Journal of Chemical Physics}\ }\textbf {\bibinfo
  {volume} {157}},\ \bibinfo {pages} {031104} (\bibinfo {year}
  {2022})}\BibitemShut {NoStop}%
\bibitem [{\citenamefont {Bradbury}\ \emph {et~al.}(2023)\citenamefont
  {Bradbury}, \citenamefont {Allen}, \citenamefont {Nguyen}, \citenamefont
  {Ibrahim},\ and\ \citenamefont {Neuhauser}}]{bradbury_bse_2023}%
  \BibitemOpen
  \bibfield  {author} {\bibinfo {author} {\bibfnamefont {N.~C.}\ \bibnamefont
  {Bradbury}}, \bibinfo {author} {\bibfnamefont {T.}~\bibnamefont {Allen}},
  \bibinfo {author} {\bibfnamefont {M.}~\bibnamefont {Nguyen}}, \bibinfo
  {author} {\bibfnamefont {K.~Z.}\ \bibnamefont {Ibrahim}},\ and\ \bibinfo
  {author} {\bibfnamefont {D.}~\bibnamefont {Neuhauser}},\ }\bibfield  {title}
  {\bibinfo {title} {{Optimized attenuated interaction: Enabling stochastic
  Bethe–Salpeter spectra for large systems}},\ }\href
  {https://doi.org/10.1063/5.0146555} {\bibfield  {journal} {\bibinfo
  {journal} {The Journal of Chemical Physics}\ }\textbf {\bibinfo {volume}
  {158}},\ \bibinfo {pages} {154104} (\bibinfo {year} {2023})}\BibitemShut
  {NoStop}%
\bibitem [{\citenamefont {Rom}\ \emph {et~al.}(1997)\citenamefont {Rom},
  \citenamefont {Charutz},\ and\ \citenamefont {Neuhauser}}]{Rom1997Neuhauser}%
  \BibitemOpen
  \bibfield  {author} {\bibinfo {author} {\bibfnamefont {N.}~\bibnamefont
  {Rom}}, \bibinfo {author} {\bibfnamefont {D.}~\bibnamefont {Charutz}},\ and\
  \bibinfo {author} {\bibfnamefont {D.}~\bibnamefont {Neuhauser}},\ }\bibfield
  {title} {\bibinfo {title} {Shifted-contour auxiliary-field monte carlo:
  circumventing the sign difficulty for electronic-structure calculations},\
  }\href {https://doi.org/10.1016/s0009-2614(97)00370-9} {\bibfield  {journal}
  {\bibinfo  {journal} {Chemical Physics Letters}\ }\textbf {\bibinfo {volume}
  {270}},\ \bibinfo {pages} {382} (\bibinfo {year} {1997})}\BibitemShut
  {NoStop}%
\bibitem [{\citenamefont {Carlson}\ \emph {et~al.}(1999)\citenamefont
  {Carlson}, \citenamefont {Gubernatis}, \citenamefont {Ortiz},\ and\
  \citenamefont {Zhang}}]{Carlson1999AFQMC}%
  \BibitemOpen
  \bibfield  {author} {\bibinfo {author} {\bibfnamefont {J.}~\bibnamefont
  {Carlson}}, \bibinfo {author} {\bibfnamefont {J.~E.}\ \bibnamefont
  {Gubernatis}}, \bibinfo {author} {\bibfnamefont {G.}~\bibnamefont {Ortiz}},\
  and\ \bibinfo {author} {\bibfnamefont {S.}~\bibnamefont {Zhang}},\ }\bibfield
   {title} {\bibinfo {title} {Issues and observations on applications of the
  constrained-path monte carlo method to many-fermion systems},\ }\href
  {https://doi.org/10.1103/physrevb.59.12788} {\bibfield  {journal} {\bibinfo
  {journal} {Physical Review B}\ }\textbf {\bibinfo {volume} {59}},\ \bibinfo
  {pages} {12788} (\bibinfo {year} {1999})}\BibitemShut {NoStop}%
\bibitem [{\citenamefont {Zhang}(2018)}]{Zhang2018AFQMC}%
  \BibitemOpen
  \bibfield  {author} {\bibinfo {author} {\bibfnamefont {S.}~\bibnamefont
  {Zhang}},\ }\bibfield  {title} {\bibinfo {title} {Ab initio electronic
  structure calculations by auxiliary-field quantum monte carlo},\ }in\ \href
  {https://doi.org/10.1007/978-3-319-42913-7_47-1} {\emph {\bibinfo {booktitle}
  {Handbook of Materials Modeling}}}\ (\bibinfo  {publisher} {Springer
  International Publishing},\ \bibinfo {year} {2018})\ pp.\ \bibinfo {pages}
  {1--27}\BibitemShut {NoStop}%
\bibitem [{\citenamefont {Baer}\ \emph {et~al.}(2013)\citenamefont {Baer},
  \citenamefont {Neuhauser},\ and\ \citenamefont
  {Rabani}}]{baer2013selfaveraging}%
  \BibitemOpen
  \bibfield  {author} {\bibinfo {author} {\bibfnamefont {R.}~\bibnamefont
  {Baer}}, \bibinfo {author} {\bibfnamefont {D.}~\bibnamefont {Neuhauser}},\
  and\ \bibinfo {author} {\bibfnamefont {E.}~\bibnamefont {Rabani}},\
  }\bibfield  {title} {\bibinfo {title} {Self-averaging stochastic kohn-sham
  density-functional theory},\ }\href
  {https://doi.org/10.1103/PhysRevLett.111.106402} {\bibfield  {journal}
  {\bibinfo  {journal} {Phys. Rev. Lett.}\ }\textbf {\bibinfo {volume} {111}},\
  \bibinfo {pages} {106402} (\bibinfo {year} {2013})}\BibitemShut {NoStop}%
\bibitem [{\citenamefont {Neuhauser}\ \emph
  {et~al.}(2014{\natexlab{b}})\citenamefont {Neuhauser}, \citenamefont {Baer},\
  and\ \citenamefont {Rabani}}]{neuhauser2014fragment}%
  \BibitemOpen
  \bibfield  {author} {\bibinfo {author} {\bibfnamefont {D.}~\bibnamefont
  {Neuhauser}}, \bibinfo {author} {\bibfnamefont {R.}~\bibnamefont {Baer}},\
  and\ \bibinfo {author} {\bibfnamefont {E.}~\bibnamefont {Rabani}},\
  }\bibfield  {title} {\bibinfo {title} {Communication: {Embedded} fragment
  stochastic density functional theory},\ }\href@noop {} {\bibfield  {journal}
  {\bibinfo  {journal} {J. Chem. Phys.}\ }\textbf {\bibinfo {volume} {141}},\
  \bibinfo {pages} {041102} (\bibinfo {year} {2014}{\natexlab{b}})}\BibitemShut
  {NoStop}%
\bibitem [{\citenamefont {Wall}\ and\ \citenamefont
  {Neuhauser}(1995)}]{Wall1995_fdg}%
  \BibitemOpen
  \bibfield  {author} {\bibinfo {author} {\bibfnamefont {M.~R.}\ \bibnamefont
  {Wall}}\ and\ \bibinfo {author} {\bibfnamefont {D.}~\bibnamefont
  {Neuhauser}},\ }\bibfield  {title} {\bibinfo {title} {Extraction, through
  filter-diagonalization, of general quantum eigenvalues or classical normal
  mode
  frequencies{\hspace{0.167em}}from{\hspace{0.167em}}a{\hspace{0.167em}}small{\hspace{0.167em}}number{\hspace{0.167em}}of{\hspace{0.167em}}residues
  or a short-time segment of a signal. i. theory and application to a
  quantum-dynamics model},\ }\href {https://doi.org/10.1063/1.468999}
  {\bibfield  {journal} {\bibinfo  {journal} {J. Chem. Phys.}\ }\textbf
  {\bibinfo {volume} {102}},\ \bibinfo {pages} {8011} (\bibinfo {year}
  {1995})}\BibitemShut {NoStop}%
\end{thebibliography}%

\end{document}